%% file: main.tex
\documentclass[12pt]{article}
\usepackage[utf8]{inputenc}
\usepackage{parskip, graphicx}
\usepackage[margin=1in]{geometry}
\input{Preamble/preamble}

\title{AdaAnn: Adaptive Annealing Scheduler for \\ Probability Density Approximation}
\author[]{Emma R. Cobian, Jonathan D. Hauenstein,
\\ Fang Liu, and Daniele E. Schiavazzi}
\affil[]{Department of Applied and Computational Mathematics and Statistics, University of Notre Dame, Notre Dame, IN, USA}
\date{}

\begin{document}

\maketitle

\begin{abstract}
\noindent Approximating probability distributions can be a challenging task, particularly when they are supported over regions of high geometrical complexity or exhibit multiple modes.
Annealing can be used to facilitate this task which is often combined with constant {\em a priori} selected increments in inverse temperature. However, using constant increments limit the computational efficiency due to the inability to adapt to situations where smooth changes in the annealed density could be handled equally well with larger increments.
We introduce AdaAnn, an adaptive annealing scheduler that automatically adjusts the temperature increments based on the expected change in the Kullback-Leibler divergence between two distributions with a sufficiently close annealing temperature. 
AdaAnn is easy to implement and can be integrated into existing sampling approaches such as normalizing flows for variational inference and Markov chain Monte Carlo.
We demonstrate the computational efficiency of the AdaAnn scheduler for variational inference with normalizing flows on a number of examples, including density approximation and parameter estimation for dynamical systems.
\end{abstract}

\section{Introduction}\label{sec:Intro}

One of most fundamental challenges in statistics and machine learning is the ability to learn a posterior distributions from its pointwise evaluations.
In this context, Markov chain Monte Carlo (MCMC) sampling is a popular paradigm to provide empirical approximations of distributions and has given rise to a large family of sampling procedures such as the Metropolis Hasting algorithm~\cite{hastings1970monte,metropolis1953equation}, the Gibbs sampler~\cite{gelfand1990sampling}, and slice sampling~\cite{neal2003slice}, among others. However, MCMC can be computationally expensive and may fail to capture complicated posterior distributions, leading to poor approximations.

Recently, optimization-based approaches using variational inference (VI)~\cite{Bishop06,Blei2017,jordan1999introduction,wainwright2008graphical} have emerged
which aim to provide a more efficient alternative to sampling-based methods with the ability to support distributions with complex shapes such as multi-modality in high-dimensional settings~\cite{ranganath2014black}.
More recent, VI approaches based on normalizing flows (NFs)~\cite{rezende2016variational}, a type of generative model, are able to characterize even complex dependence in multivariate distributions. 
They offer a flexible framework by transforming a base distribution through a composition of invertible mappings until the desired complexity has been attained.

There are many different types of NFs such as planar flows~\cite{rezende2016variational}, radial flows~\cite{rezende2016variational}, realNVP~\cite{dinh2016density}, autogressive flows that include inverse autoregressive flow (IAF)~\cite{kingma2016improved} and masked autoregressive Flow (MAF)~\cite{papamakarios2018masked}, and glow~\cite{kingma2018glow}, among others.
An introduction to the fundamental principles of NFs including their expressive power and computational trade-offs, together with a review of a wide verity of flow formulations are provided in~\cite{Kobyzev_2020, papamakarios2018masked}.
They have been applied in various settings for VI such as 
density estimation and sampling since their introduction. 
For example, NFs are used to formulate Gaussian processes as function priors~\cite{maronas2021transforming}
while \cite{liu2019graph} introduces NFs to the setting of graph neural networks for prediction and generation. In~\cite{yang2019pointflow}, NFs are applied to 3D point cloud generation; \cite{louizos2017multiplicative} apply NFs to approximate the latent variables in Bayesian neural networks. 
Recent applications of NFs include semi-supervised learning~\cite{izmailov2020semi}, coupling with surrogate modelling for inference with computationally expensive models~\cite{wang2021variational},  and solving inverse problems~\cite{whang2021composing}, among others.

In this study, we focus on VI via NFs, specifically on situations where the target distribution to be approximated is supported over a geometrically complex subset of the parameter space or has multiple modes.
Rather than designing new types of NFs offering improved representations of multimodal densities, we choose instead to approximate a collection of intermediate smoother posteriors generated through a parameterization defined in terms of an annealing temperature.

Annealing or tempering of probability density functions is used in optimization (e.g.,  simulated annealing~\cite{kirkpatrick1983optimization} and simulated tempering~\cite{marinari1992simulated}) and MCMC sampling to generate realizations from complex and multimodal distributions (e.g. tempered transition~\cite{Neal1996} and parallel tempering~\cite{geyer1991markov}). 
Tempering is also used in Bayesian statistics to study theoretical properties and concentration rates for posterior distributions~\cite{bhattacharya2019bayesian}. 
This has been extended in~\cite{alquier2019concentration} to analyze the concentration of VI approximations of (tempered) posteriors
while \cite{huang2018improving} develops an annealed version of the objective functions in VI to improve inferential explorability. 
In the setting of NFs, \cite{rezende2016variational} applies an annealed version of the free energy for VI via NFs by approximating a series of tempered distributions with slowly decreased temperature to provide better results on the final approximated target distribution.

Various temperature cooling schedules have been proposed to improve computational efficiency in simulated annealing such as simple linear schedules~\cite{kirkpatrick1983optimization}, exponential multiplicative cooling~\cite{kirkpatrick1983optimization}, and 
logarithmic multiplicative cooling~\cite{aarts1989boltzmann}, among others.
There also exists work on adaptive cooling where the temperature at each state transition depends on an adaptive factor based on the difference between the current solution and the best achieved solution of an objective function, including some recent work~\cite{karabin2020simulated,mahdi2017performance}.
Outside the realm of simulated annealing, annealing strategies and cooling schedules have received little~attention. 

We use a simple instance of NFs, namely, planar flows~\cite{rezende2016variational}, to motivate our methodological development for an annealing scheduler in the settings of VI via NFs. 
Planar flows are shown to be a universal approximator in $L_1$ for one-dimensional problems in theory~\cite{kong2020expressive}, but have been sometimes associated with a limited approximation power and more complex flow formulations have often been preferred in applications, limiting the analysis of this flow in the literature, particularly for higher dimensional latent spaces and complicated posterior distributions.
We outline cases where planar flows alone fail to capture the structure of a multimodal density but
the combination with annealing leads to successful approximations.

Our main contribution is \emph{AdaAnn} (Adaptive Annealing), a novel scheduler that adaptively selects the change in temperature during the annealing process by tracking the Kullback-Leibler divergence between successive temperature changes. Through five examples, we demonstrate that AdaAnn helps NFs converge to the target posterior and leads to significant computational savings compared to a linear scheduler for both univariate and multivariate cases. 
In addition, we show how planar flows with AdaAnn achieve better approximation to the target distribution compared to more expressive flows but without~using~annealing.

The remainder of the paper is organized as follows. Section~\ref{sec:Background} provides necessary background information regarding NFs and VI. Section~\ref{sec:Methods} describes AdaAnn, our new adaptive annealing schedule for VI via NFs. 
Five examples are presented in Section~\ref{sec:Examples} which demonstrate the superior performance of using annealing for VI via NFs, and the computational advantage of AdaAnn over linear annealing schedulers.
We conclude with a discussion in Section~\ref{sec:discussion}.

\section{Background}\label{sec:Background}

\subsection{Normalizing Flows}\label{sec:NormFlow}

Normalizing flows are compositions of invertible and differentiable mappings used to transform samples from a base probability density function (pdf) $q_0$, e.g., a standard Gaussian, into samples from a desired distribution and vice-versa.
Consider a single layer of a normalizing flow 
with a bijection $f: \mathbb{R}^d \rightarrow \mathbb{R}^d$ that maps a set of $N$ sample points~$\{\bm{z}^{(i)}_{0}\}_{i=1}^{N}$ where $\bm{z}^{(i)}_{0}\sim \bm{Z}_{0},\,i=1,\dots,N$, 
from the base density to $\{\bm{z}^{(i)}_{1}\}_{i=1}^{N}$ where$\bm{z}^{(i)}_{1} = f(\bm{z}^{(i)}_{0}),\,i=1,\dots,N$,
and $d$ is the dimension 
of $\bm Z_{0}$ and $\bm Z_{1}$.
Given $\bm{Z}_{0}\sim q_0$, the density of the transformed variables~$\bm{Z}_{1}\sim q_1$ can be computed using the change of variables formula and the properties of inverse functions, namely
\begin{equation}\label{eq:q1}
    q_1(\bm{Z}_1) = q_0(f^{-1}(\bm{Z}_1))\cdot \Big| \det \Big( \pdv{f^{-1}}{\bm{Z}_1} \Big) \Big| = q_0(\bm{Z}_0)\cdot \Big|\det \Big( \pdv{f}{\bm{Z}_0} \Big) \Big|^{-1}.
\end{equation}

One can easily generalize this to $L$ layers
of transformations
so that the initial set of sample points are transformed to
\begin{equation}\label{eq:zk composed}
    \bm{z}^{(i)}_{L} = f_L \circ f_{L-1} \circ \cdots \circ f_2 \circ f_1(\bm{z}^{(i)}_{0}),\,\,\,i=1,\dots,N,
\end{equation}
and the corresponding pdf is given by
\begin{equation}\label{eq:qk}
    q_L(\bm{Z}_L) = q_0(\bm{Z}_0)\cdot \prod_{\ell=1}^{L} \Big|\det \Big( \pdv{f_\ell}{\bm{Z}_{\ell-1}} \Big) \Big|^{-1}.
\end{equation}
To simplify the computation, a desirable property of flow $f_{\ell}$ is that
the Jacobian determinant is easy to compute, e.g., through the product of the diagonal entries, as in lower triangular Jacobian matrices.
Many different formulations of NFs have been investigated in the literature. In this paper, we use planar flows and the real-valued Non Volume Preserving (realNVP) flows to demonstrate our proposed methodology, which are summarized next.

Planar flows \cite{rezende2016variational} are one of the simpler instances of NFs where each layer transforms a set of samples with expansions or contractions perpendicular to a $d$-dimensional hyperplane. 
A planar flow $f: \mathbb{R}^d \cross \mathbb{R}^{2d+1} \rightarrow \mathbb{R}^d$ consists of an activation function $h:\mathbb{R}\rightarrow\mathbb{R}$
and parameters $\phi = \{\bm{u} \in \mathbb{R}^d, \bm{w} \in \mathbb{R}^d, b \in \mathbb{R}\}$ such that:
\begin{equation}\label{eq:planar flow}
    f(\bm{Z}; \phi) = \bm{Z} + \bm{u}\cdot h(\bm{w}^T \bm{Z} + b).
\end{equation}
When $\bm{u}^T \bm{w} \geq -1$, this flow is invertibile~\cite{rezende2016variational} 
and its Jacobian determinant is equal to
\begin{equation}\label{eq:det}
    \Big|\det \Big(\pdv{f}{\bm{Z}} \Big) \Big| = |\det (\bm{I} + \bm{u}\,(\bm{w} h'(\bm{w}^T\,\bm{Z} + b))^T)| = |1 + \bm{u}^T \,\bm{w} h'(\bm{w}^T\,\bm{Z} + b)|,
\end{equation}
where $h'$ is the derivative of $h$.  
With $L$ layers, the transformed random variable
\begin{equation}
    \bm{Z}_L = f_L(\bullet ; \phi_L) \circ f_{L-1}(\bullet; \phi_{L-1}) \circ \cdots \circ f_2(\bullet; \phi_2) \circ f_1(\bm{Z}_0; \phi_1)
\end{equation}
has corresponding pdf
\begin{equation}\label{eq:qk planar flow}
    q_L(\bm{Z}_L) = q_0(\bm{Z}_0)\prod_{\ell=1}^L |1 + \bm{u}_\ell^T \bm{w}_\ell \cdot h'(\bm{w}_\ell^T \bm{Z}_{\ell-1} + b_\ell)|^{-1}.
\end{equation}

To enhance the expressiveness of NFs while maintaining a linear complexity in the computation of the Jacobian determinant, dependencies between different components of latent vectors $\bm{Z}_{\ell},\,\ell=1,\dots,L$, can be introduced through autoregressive transformations.
A widely used auto-regressive flow is realNVP,  defined as
\begin{equation}\label{equ:dirRNVP}
\!\!\!\!Z_{\ell+1,j}\!=\!
\begin{cases}
Z_{\ell,j}, & \text{for}\,\,j = 1,\dots,c,\\
Z_{\ell,j} \exp(a_{s_k}(Z_{\ell,1},\dots,Z_{\ell,c}))\!+\! a_{t_k}(Z_{\ell,1},\dots,Z_{\ell,c}) & \text{for}\,\,j\!=\!c\!+\!1,\dots,d,\,k\!=\!j\!-\!c,
\end{cases}\!\!
\end{equation}
where $Z_{\ell+1,j}$ denotes the $j^{\rm th}$ component of $\bm{Z}_{\ell+1}$ in layer $\ell+1$, and  $a_{s_k}$ and $a_{t_k}$ are scale and translation functions in layer $k$, respectively, 
and are usually implemented as neural networks.
The components in $\bm{Z}$ are divided into two groups in Eq.~\eqref{equ:dirRNVP}. The variables in the first group are copied directly into the next layer whereas the remaining variables go through an autoregressive transformation. The roles of the two groups are reversed (or the variables are randomly scrambled) after every layer.
Since the $c^{\rm th}$ component of $\bm{Z}_{\ell+1}$ in layer $\ell+1$ depends only on the components $1,\dots,c$ of $\bm{Z}_{\ell}$, the Jacobian matrix is lower triangular and its determinant is simply the product of the diagonal entries $\prod_{k=1}^{d-c} a_k(\bm{Z}_{k-1})$.
In particular, realNVP is efficient and has the same computational complexity for
sampling and density estimation~\cite{dinh2016density}. 
Even if the mappings $\bm{a}_s$ and $\bm{a}_t$ are not invertible, the transformation in Eq.~\eqref{equ:dirRNVP} is still invertible since
\begin{equation}\label{equ:invRNVP}
\!\!\!\!Z_{\ell,j}\!= \!
\begin{cases}
Z_{\ell+1,i}, &\!\!\text{for}\,\,j=1,\dots,c,\\
[Z_{\ell+1,j}\!-\!a_{t_k}\!(Z_{\ell,1},\dots,Z_{\ell,c})]\exp (- a_{s_k}\!(Z_{\ell,1},\dots,Z_{\ell,c})) &\!\!\! \text{for}\,\,j\!=\!c+1,\dots,d,\,k\!=\!j\!-\!c.
\end{cases}\!\!
\end{equation}

\subsection{Variational Inference via Normalizing Flows}\label{sec:loss functions}

Variational inference is a common method for statistical inference and machine learning that approximates probability densities by minimizing their Kullback-Leibler (KL) divergence from a target distribution.
In particular, VI provides an effective alternative to sampling-based approaches for density approximation such as MCMC. It is based on optimization and designed to offer improved computational efficiency. Additionally, one of the major applications of NFs is VI.
Without loss of generality, we illustrate the application of NFs for VI in approximating the posterior distribution $p(\bm{Z}|\bm{X})$ of the model parameters $\bm{Z}$ given observed data $\bm{X}$. Such an approximation is obtained by minimizing the free energy $\mathcal{F}$, the negative of which is a lower bound to the marginal log-density function $\log p(\bm{X})$ (a.k.a., the evidence). 
Due to the analytical difficulty in maximizing the marginal log-density function, the minimization of the free energy is often used in VI. 
If $q_{\phi}(\bm{Z}\vert\bm{X})$ is the variational distribution with parameters $\phi$ that approximates the true posterior $p(\bm{Z}|\bm{X})$, the free energy~is
\begin{equation}\label{equ:elbo}
\begin{split}
\mathcal{F}(\bm{X},\phi) &= \mathbb{D}[q_{\phi}(\bm{Z}\vert\bm{X})\,\Vert\,p(\bm{Z})] - \mathbb{E}_{q_{\boldsymbol{\phi}}}[\log\,p(\bm{X}\vert \bm{Z})] \\
&= \mathbb{E}_{q_{\boldsymbol{\phi}}}\left[\log q_{\boldsymbol{\phi}}(\bm{Z}\vert\bm{X}) - \log p(\bm Z,\bm X)\right]
\end{split}
\end{equation}
where $\mathbb{D}[\cdot\Vert\cdot]$ denotes the KL divergence between two distributions.
Following the notation in Section~\ref{sec:NormFlow}, we express the density $q_{\phi}(\bm{Z}\vert\bm{X})$ as $q_{L}(\bm{Z}_{L})$ and apply the change of variables formula in Eq.~\eqref{eq:qk planar flow} to~Eq.~\eqref{equ:elbo} to obtain
\begin{equation}\label{eq:FEB}
\begin{aligned}
    \mathcal{F}(\bm{X},\phi)&=\mathbb{E}_{q_0}\left[\log q_L(\bm{Z}_L) - \log p(\bm X,\bm Z_L)\right]\\
    &=\mathbb{E}_{q_0}[\log q_0(\bm{Z}_0)]-
    \mathbb{E}_{q_0}\left[\sum_{\ell=1}^L\log\bigg|\det\frac{\partial f_\ell}{\partial \bm{z}_{\ell-1}}\bigg|\right]+\mathbb{E}_{q_0}[\log(p(\bm X,\bm Z_L))].
\end{aligned}
\end{equation}
Minimization of the free energy $\mathcal{F}$ with respect to the parameters $\phi$ 
is often achieved through gradient-based optimization, e.g., stochastic gradient descent, RMSprop \cite{hinton2012}, Adam \cite{kingma2017adam}, and others. 
The expectations in Eq.~\eqref{eq:FEB} are often replaced by their Monte Carlo estimates
by using~$N$ realizations from the base distribution $q_0$.
For planar flows, Eq.~\eqref{eq:FEB} becomes
\begin{equation}\label{eq:KL}
\mathcal{F}(\bm{X},\phi)\!\approx\!\frac{1}{N}\! \sum_{i=1}^{N}\! \Big[\! \log\!(q_0(\bm{z}_{0,i}))\!-\!\log(p(\bm{z}_{L,i},\bm{X}))\!-\! \sum_{\ell=1}^{L}\! \log\! \abs{1\! +\! \bm{u}_\ell^T \bm{w}_{\ell}\,h(\bm{w}_\ell^T \bm{z}_{\ell-1,i}\! +\! b_\ell)} \Big].\!
\end{equation}

\subsection{Annealing}\label{sec:annealing}

Annealing is a useful technique when sampling from complicated distributions. Coupled with MCMC techniques or VI, annealing can help improve sampling efficiency and accuracy.  During the application of annealing, the annealing temperature $1/t$ continuously decreases~with
\begin{equation}\label{eq:continuousAnnealing}
p_t(\bm{Z},\bm{X}) = p^t(\bm{Z},\bm{X}),\,\,\text{for } t \in (0,1].
\end{equation}
In practice, a discrete version of Eq.~\eqref{eq:continuousAnnealing} is used by generating a sequence of functions
\begin{equation}
p_k(\bm{Z},\bm{X}) = p^{t_k}(\bm{Z},\bm{X}),\,\,\text{for } k=0,\dots,K
\end{equation}
where $0 < t_{0} < \cdots < t_{K} \le 1$ is an annealing scheduler and $p_{k}(\bm{Z},\bm{X})$ is the annealed or tempered distribution.  A commonly used annealing schedule is linear \cite{rezende2016variational} of the form
\mbox{$t_j=t_{0} + j\dot (1-t_{0})/K$} for \mbox{$j=0,\ldots,K$}
with constant increments 
\mbox{$\epsilon = (1-t_{0})/K$}. 
For example, when combining annealing with VI and planar flows, the free energy $\mathcal{F}$ in Eq.~\eqref{eq:KL}~is
\begin{equation}\label{eq:KLannealing}
\!\!\!\mathcal{F}(\bm{x},\phi)\!\approx\!\frac{1}{N}\!\sum_{i=1}^{N}\!\Big[\!\log (q_0(\bm{z}_{0,i}))\! - t_k \log(p(\bm{z}_{L,i},\bm{X})) \!-\! \sum_{\ell=1}^{L} \!\log \abs{1\! + \!\bm{u}_\ell^T \bm{w}_\ell h(\bm{w}_\ell^T \bm{z}_{\ell-1,i} \!+\! b_\ell)} \Big].\!
\end{equation}

\subsection{A Motivating Example}\label{sec:meg}

\begin{figure}[!htb]
    \centering
    $\begin{array}{cc}
    \includegraphics{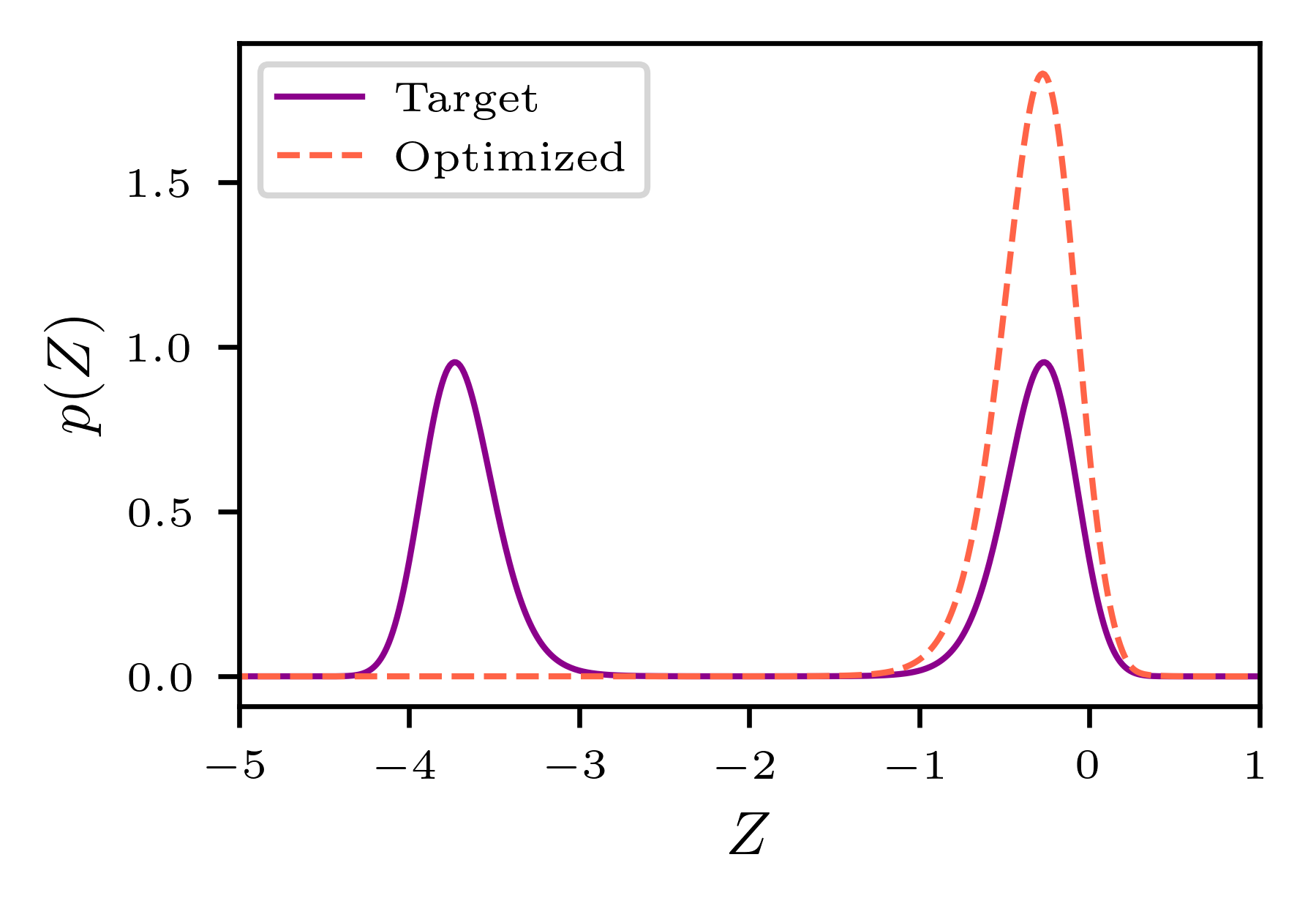} &
    \includegraphics{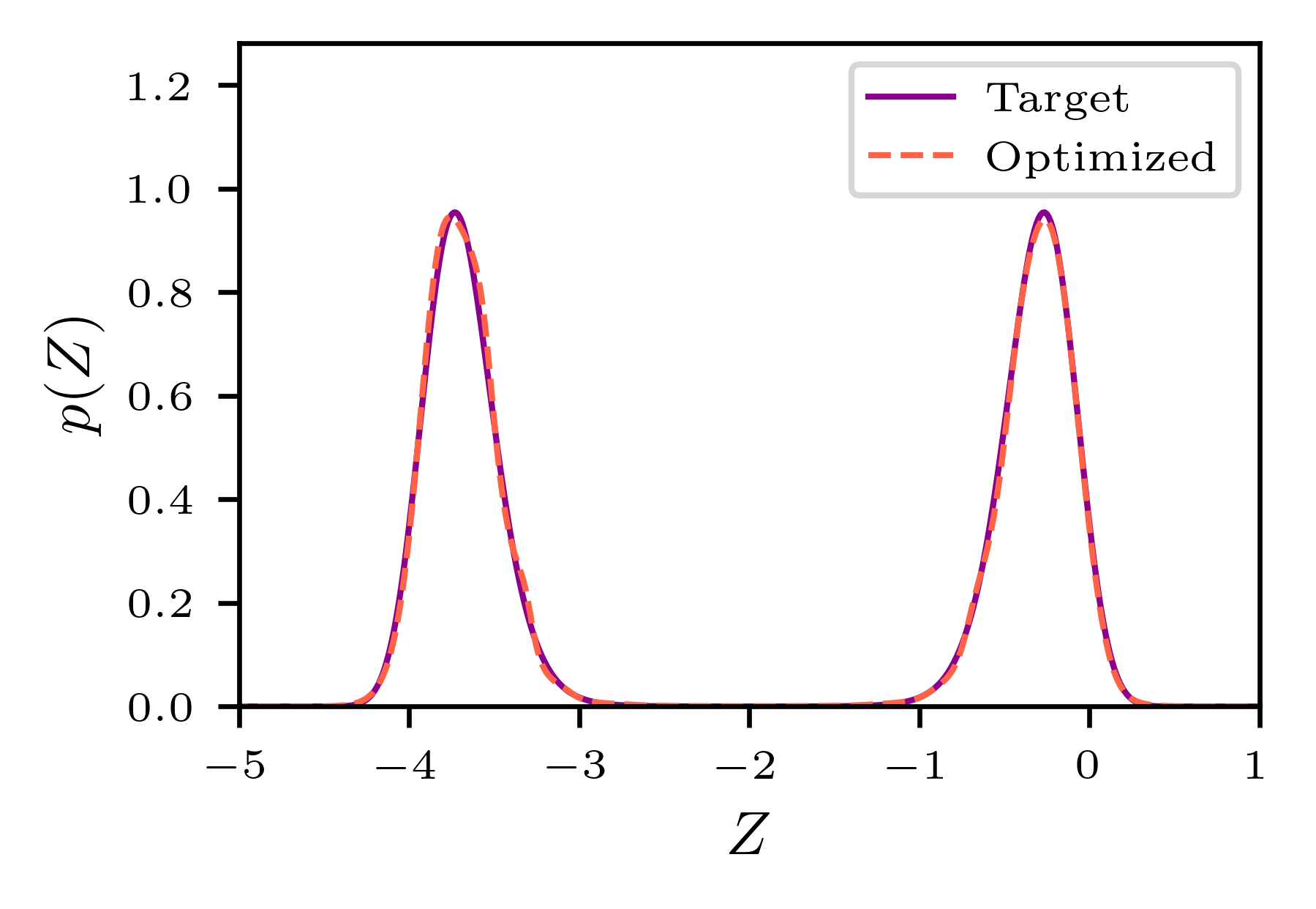}\\[-10pt]
    \hbox{(a) No Annealing} & \hbox{(b) Linear Annealing Scheduler}\end{array}$
    \caption{Variational approximation for bimodal density $p(Z)$ with and without annealing.} \label{fig:bimodal_1mode}  \vspace{-12pt}
\end{figure}

Consider sampling from the pdf $p:\mathbb{R} \rightarrow \mathbb{R}$ where
\begin{equation}\label{eq:bimodal}
p(Z) = 0.954 \cdot e^{-[(Z+2)^2 - 3]^2}.
\end{equation}
Hence, $p(Z)$ is a bimodal distribution with peaks at $Z=-2\pm\sqrt{3}$ 
which is the ``target'' in Figure~\ref{fig:bimodal_1mode}.
Consider its variational approximation $q_L(Z) \approx p(Z)$ obtained by transforming a base distribution \mbox{$N(\mu=0,\sigma^2=4)$} using a composition of $L=50$ planar flow layers with hyperbolic tangent activation.
We use Adam optimizer with a learning rate of 0.005 and train 8,000 iterations consisting of $N=100$ sample points each.
In our experiment, the outcome yields Figure~\ref{fig:bimodal_1mode}(a) 
which suggests that the optimal $q_L$ without annealing is only able to capture a single mode.
Using the same planar flow but with annealing as given in Eq.~\eqref{eq:KLannealing}, 
our experiment showed that both modes were captured as shown in Figure~\ref{fig:bimodal_1mode}(b).

The annealing strategy used in Figure~\ref{fig:bimodal_1mode}(b) had a schedule with an initial inverse temperature of $t_0=0.01$ that increases with a constant step size of $\epsilon=10^{-4}$. 
The Adam optimizer ran for $500$ iterations at $t_0$\footnote{since the basic distribution may be significantly different than the first annealed distribution, we used a larger number of iterations at $t_0$.} and one iteration afterwards throughout the annealing process. 
In addition,  8,000 additional iterations were run when the temperature reaches $t=1$, indicated as the \emph{refinement training} phase.\footnote{This phase allows for a more refined approximation to $p(Z)$ through an increased number of iterations and sample points.}  All together, a total of 10,400 iterations were run through NFs before the annealing temperature reached 1. 

This relatively large number of iterations is rather typical with linear annealing schedulers to reach a variational approximation of a target distribution with satisfactory accuracy. 
The large number of iterations is due to the typical small steps of constant size characterizing linear annealing schedulers (e.g., $10^{-4}$ in the above example). 
An exceedingly large temperature step during the annealing process could lead to a sub-optimal approximation that does not capture the main structural features of the target distribution (e.g., missing a mode in a multi-modal distribution).


In the following, we propose a new annealing strategy that can significantly cut down the number of iterations in NFs for VI without sacrificing the quality of the final approximation.

\section{Method}\label{sec:Methods}
The following proposes the AdaAnn scheduler, a new \emph{adaptive annealing scheduler}, that uses an adjustable step size $\epsilon_{k} = \epsilon_{k}(t) > 0, k=1,\dots,K$, designed to reduce the number
of steps $K$ as much as possible while providing accurate distributional approximations in VI via NFs.

\subsection{AdaAnn Scheduler}
Intuitively, small temperature changes are desirable to carefully explore the parameter spaces at the beginning of the annealing process, whereas larger changes can be taken as $t_{k}$ increases after annealing has helped the approximate distribution to capture important features of the target distribution (e.g., locating all the relevant modes).
In VI, the KL-based loss function in Eq.~\eqref{equ:elbo} can be used as a metric to adjust the annealing temperature increment.
In this context, the proposed AdaAnn scheduler determines the increment $\epsilon_{k}$ that approximately produces a pre-defined change in the KL divergence between two distributions tempered 
at~$t_{k}$ and $t_{k+1}=t_{k}+\epsilon_{k}$, respectively.
%
%
In particular, the KL divergence between these two distributions is given by
\begin{equation}\label{equ:kl_exp}
 \mathbb{D}[p^{t_{k}}(\bm Z)||p^{t_{k}+\epsilon_{k}}(\bm Z)] = 
 \displaystyle
 \int c(t_{k})\,p^{t_{k}}(\bm Z) \log \left( \frac{c(t_{k})\, p^{t_{k}}(\bm Z)}{c(t_{k}+\epsilon_k)\,p^{t_{k}+\epsilon_k}(\bm Z)} \right) d\bm Z,
\end{equation}
where $c(s) = 1/\int p^{s}(\bm Z)\,d\bm Z$ denotes the normalizing constant associated with $p^{s}(\bm Z)$.
A Taylor series expansion of the right hand side of Eq.~\eqref{equ:kl_exp} leads to the following.

\begin{theorem}\label{thrm:adaAnnE}
For two tempered pdfs~$p^{t_{k}}$ and $p^{t_{k}+\epsilon_{k}}$ with annealing step $\epsilon_{k}$, the KL divergence~is
    \begin{equation}\label{eqn:KL}
    \mathbb{D}[p^{t_{k}}(\bm Z)\Vert p^{t_{k}+\epsilon_{k}}(\bm Z)] = \frac{\epsilon_k^2}{2}\,\mathbb{V}_{p^{t_{k}}} [\log p(\bm Z)] + O(\epsilon_{k}^3) \approx \frac{\epsilon_{k}^2}{2} \mathbb{V}_{p^{t_{k}}} [\log p(\bm Z)].
    \end{equation}
    Letting the KL divergence equal a constant $\tau^2/2$, where $\tau$ is referred to as the KL divergence tolerance, the step size $\epsilon_k$ becomes
    \begin{equation}\label{equ:adaptive_rule}
    \epsilon_k = \frac{\tau}{\sqrt{\mathbb{V}_{p^{t_k}}[\log p(\bm Z)]}}.
    \end{equation}
    
    \begin{proof}
    For simplifying the presentation, we avoid using subscripts.
        From the definition of KL divergence, we have
        \begin{equation*}
        \begin{split}
            \mathbb{D}[p^t(\bm Z)||p^{t+\epsilon}(\bm Z)] &= \int c(t) \cdot p^t(\bm Z)
            \cdot \log \left( \frac{c(t) \cdot p^t(\bm Z)}{c(t+\epsilon) \cdot p^{t+\epsilon }(\bm Z)} \right) d\bm Z\\
            &= \int c(t) \cdot p^t(\bm Z) \cdot \log \left( \frac{c(t)}{c(t+\epsilon)} \cdot p^{-\epsilon} (\bm Z) \right) d\bm Z.
        \end{split}            
        \end{equation*}
        The Taylor expansion of $c(t)/ c(t+\epsilon)$ has the form
        \begin{equation*}
        \begin{aligned}
            \dfrac{c(t)}{c(t+\epsilon)} &= c(t) \int p^{t+\epsilon}(\bm Z)\,d\bm Z = c(t)  \int p^{t}(\bm Z)\left[1 + \epsilon \log p(\bm Z) + \frac{[\epsilon \log p(\bm Z)]^2}{2} + \dots \right] d\bm Z\\
           &= c(t)\!\int\! p^{t}(\bm Z)\,d\bm Z + c(t)\! \int\! p^{t}(\bm Z)\,\epsilon\,\log p(\bm Z)\,d\bm Z + c(t)\! \int\! p^{t}(\bm Z) \frac{[\epsilon \log p(\bm Z)]^2}{2}\,d\bm Z \!+ \cdots \\
          &= 1 + \epsilon\,\mathbb{E}_{p^t} [\log p(\bm Z)] + \frac{\epsilon^2}{2} \mathbb{E}_{p^t} [\log (p(\bm Z)^2)] + O(\epsilon^3)
        \end{aligned}
        \end{equation*}
        and its logarithm is
        \begin{equation*}
        \begin{aligned}
            \log \left(\frac{c(t)}{c(t+\epsilon)}\right) &= \log (1 + \epsilon \cdot \mathbb{E}_{p^t} [\log p(\bm Z)] + \frac{\epsilon^2}{2} \mathbb{E}_{p^t} [\log (p(\bm Z)^2)] + O(\epsilon^3))\\
            &= \epsilon\,\mathbb{E}_{p^t} [\log p(\bm Z)] + \frac{\epsilon^2}{2}\,\mathbb{E}_{p^t} [(\log p(\bm Z))^2] - \frac{\epsilon^2}{2}\,\mathbb{E}_{p^t}[\log p(\bm Z)]^2 + O(\epsilon^3)\\
            &= \epsilon\,\mathbb{E}_{p^t} [\log p(\bm Z)] + \frac{\epsilon^2}{2}\,\mathbb{V}_{p^t}[\log p(\bm Z)] + O(\epsilon^3).
        \end{aligned}
        \end{equation*}
        Putting everything together with $\log p^{-\epsilon}(\bm Z) = -\epsilon \log p(\bm Z)$, we have
        \begin{equation*}
        \begin{aligned}
       \mathbb{D}[p^t(\bm Z)||p^{t+\epsilon}(\bm Z)] &=\!\int\!c(t) p^t(\bm Z) \left\{\! \epsilon\, \mathbb{E}_{p^t} [\log p(\bm Z)]\! +\! \frac{\epsilon^2}{2} \mathbb{V}_{p^t}[\log p(\bm Z)]\! +\! O(\epsilon^3)\! - \!\epsilon \log p(\bm Z) \!\right\} d\bm Z\\
        &= \epsilon\,\mathbb{E}_{p^t} [\log p(\bm Z)] + \frac{\epsilon^2}{2}\,\mathbb{V}_{p^t} [\log p(\bm Z)] - \epsilon\,\mathbb{E}_{p^t} [\log p(\bm Z)]\\
        &= \frac{\epsilon^2}{2}\,\mathbb{V}_{p^t} [\log p(\bm Z)] + O(\epsilon^3).
        \end{aligned}
        \end{equation*}
    \end{proof}
\end{theorem}

The quantity $\mathbb{V}_{p^{t_{k}}}[\log p(\bm Z)]$ in Theorem \ref{thrm:adaAnnE} can be approximated using a Monte Carlo (MC) estimate with samples from $q^{t_k}_L \approx p^{t_k}$ available from NFs at a given temperature $t_{k}$.
Specifically, we draw $M$ samples, $\bm{z}^{(i)}_L,\,i=1, \dots,M$, and compute the sample variance of $\{\log p(\bm{z}^{(i)})\}_{i=1}^{M}$.
This MC approximation also provides the following 
intuitive interpretation of the AdaAnn scheduler from Theorem~\ref{thrm:adaAnnE}.

At the beginning of the annealing process, $t_0$ is small and the tempered distribution $p^{t_0}$ is rather flat, therefore samples from this distribution cover almost equally well the high density regions in the support of $p$ and its tails leading to a large variance of $\log(p)$.
The combination of a large variance of $\log(p)$ with 
the given constant $\tau$ (see Eq.~\eqref{equ:adaptive_rule}) results in a small annealing increment $\epsilon_k$. 
As $t$ increases, $p^t$ becomes closer and closer to the target $p$, leading to most of the samples from $q_L^t$ falling in high-density regions of the target $p$. 
This causes the variance of $\log(p)$ to shrink, resulting in larger increments $\epsilon_k$. 

In summary, the mathematical formulation in Eq.~\eqref{equ:adaptive_rule} reflects the sensitivity of the annealing process in capturing the shape of the target distribution. 
In particular, $t$ should increase slowly at the beginning of the annealing process due to rapid changes in the KL divergence at high temperatures, whereas the tempered distribution becomes less sensitive to temperature changes as it becomes increasingly similar to the target distribution.

Algorithm~\ref{alg:adaptive} summarizes the implementation of the AdaAnn scheduler with NFs. Source code is available
at \url{https://github.com/ercobian/AdaAnn-VI-NF}.

\begin{algorithm}[H]
\caption{AdaAnn Scheduler}\label{alg:adaptive}
\begin{algorithmic}
\State \textbf{input}: initial  temperature $t^{-1}_0$, target distribution $p$, number of iterations $T_0$ at $t_0$,  number of iterations $T_1$ at $t=1$, number of iterations $T$ for $t\in(t_0,1)$, number of NF samples $N$ for $t\in[t_0,1)$, number of NF samples $N_1$ for $t=1$, number of MC samples $M$ for calculation of $\epsilon$,  KL divergence tolerance $\tau$, a prespecified NF structure with~$L$ layers of transformation.
\State \textbf{output}: approximated distribution $q_{L}$ for $p$.
\State $t \gets t_0$;  $\epsilon \gets 0$
\While{$t+\epsilon < 1$}
    \State $t \gets t + \epsilon$
    \State Obtain an empirical approximation  $q^t$ to $p^t$ with $N$ samples with NF for the specified 
    \State \hspace{12pt} 
    number of iterations at $t$ ($T_0$ for $t= t_0$ and $T$ for $t\in (t_0,1)$);
    \State Calculate the MC estimate of $\mathbb{V}_{p^{t}}[\log p(\bm Z)]$ in Eq.~\eqref{equ:adaptive_rule} using $\bm z^{(i)}\sim q^t, i=1,\dots,M$:
    \State \hspace{12pt}  $\textstyle S^2 =(M-1)^{-1}\sum_{i=1}^M (\log p(\bm z^{(i)})- \overline{\log p(\bm z)})^2$, where $\overline{\log p(\bm z)}= M^{-1}\sum_{i=1}^M \log p(\bm z^{(i)})$;
    \State $\epsilon \gets \tau/S$
\EndWhile
\State $t \gets 1$
\State (Optional) Refine at $t=1$ by running the NFs for $T_1$ iterations to obtain a final approximation $q$ to $p$ with $N_1$ samples.
\end{algorithmic}
\end{algorithm}

\section{Numerical Examples}\label{sec:Examples}
The following summarizes applying AdaAnn to five examples: three synthetic cases and two applications in dynamical systems. 
We first compare AdaAnn with linear schedulers in one-dimensional settings with bimodal  distributions. We then examine two-dimensional bimodal densities and compare the performance of a planar flow with AdaAnn and of a flow with greater approximation power (i.e., realNVP).
For the two applications in dynamical systems, we obtain posterior variational inference of the parameters of a Lorenz attractor and a non-linear dynamical system simulating HIV viral dynamics, respectively, via NFs with AdaAnn.
For all examples, unless otherwise noted, we use hyperbolic tangent activation functions in planar flows and optimize the free energy loss function in VI via NFs using Adam.

\subsection{Example 1:  One-dimensional Bimodal Distribution}\label{sec:Comparing_Methods}
We apply AdaAnn to the bimodal density in Eq.~\eqref{eq:bimodal} and compare with the linear annealing scheduler in Section~\ref{sec:meg}. The same planar flow as specified in  Section~\ref{sec:meg} was employed. 
We use Algorithm~\ref{alg:adaptive} with the following hyperparameters: $t_0=0.01$ (identical to the linear scheduler), $T_0=500$, $T=2$, $T_1=\mbox{8,000}$, $\tau=0.01$ and $M=\mbox{1,000}$.
For the Adam optimizer, we applied the same learning rate schedule  as in Section~\ref{sec:meg}.
The number of points in each iteration increases from $N=100$ to $N_1=\mbox{1,000}$ during the refinement stage at $t=1$.

\begin{figure}[!htb]
    \centering
    \includegraphics[scale=1.1]{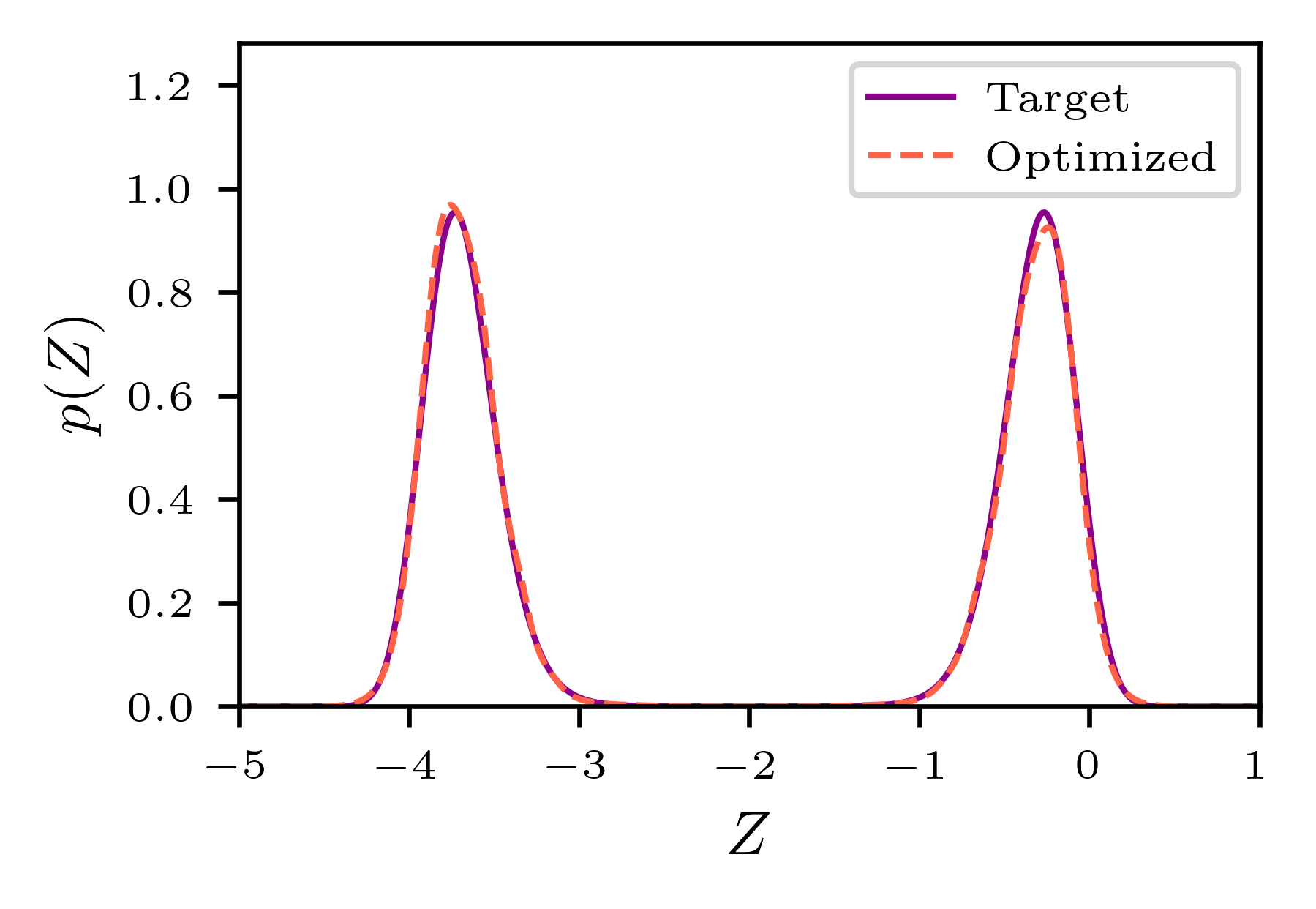}\vspace{-15pt}
    \caption{Variational approximation of $p(Z)$ in Example 1 with the AdaAnn scheduler.}
    \label{fig:bimodal_adaAnn}\vspace{-6pt}
\end{figure}

\begin{figure}[!htb]
\centering
$\begin{array}{cc}
\includegraphics{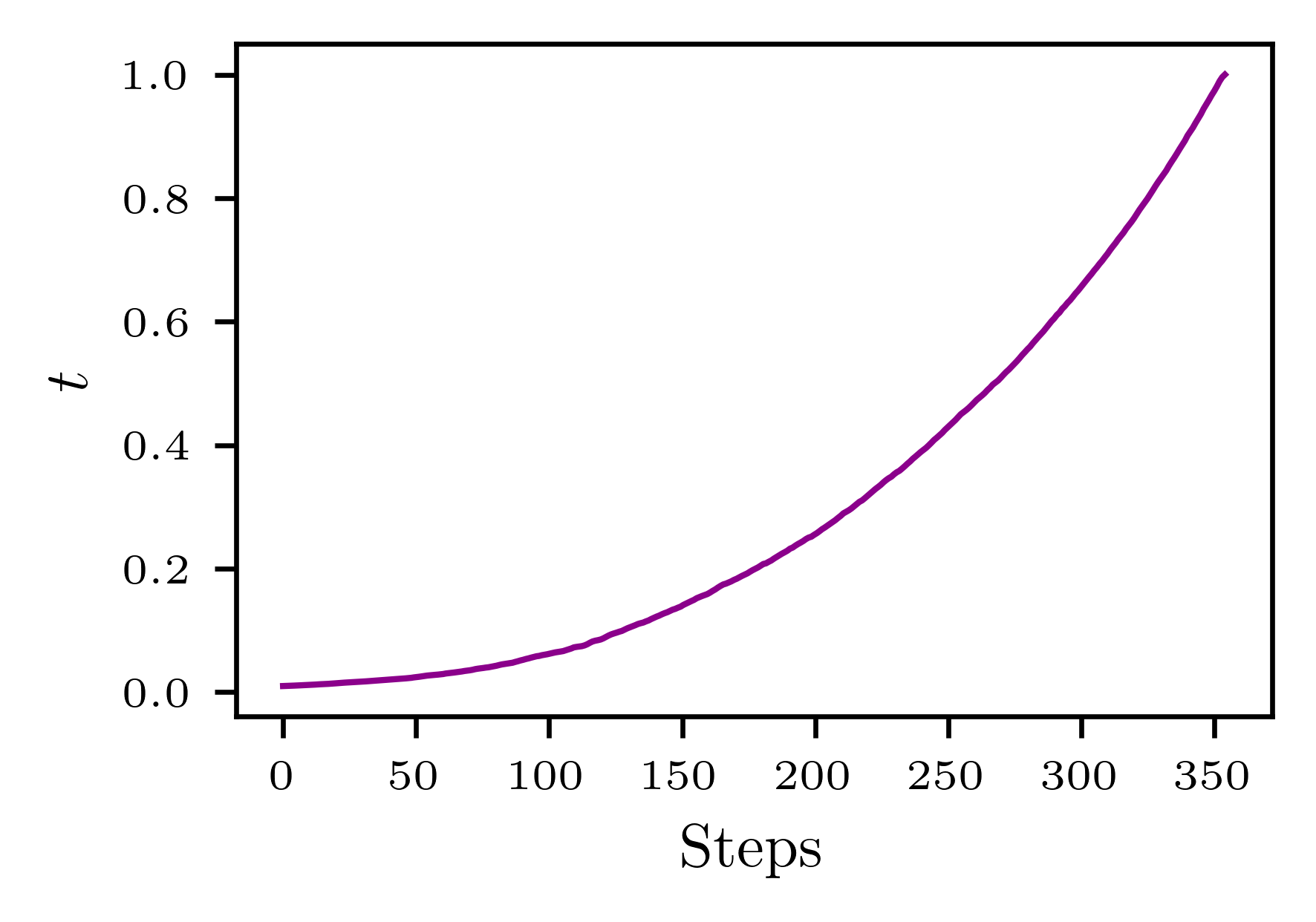} &
\includegraphics{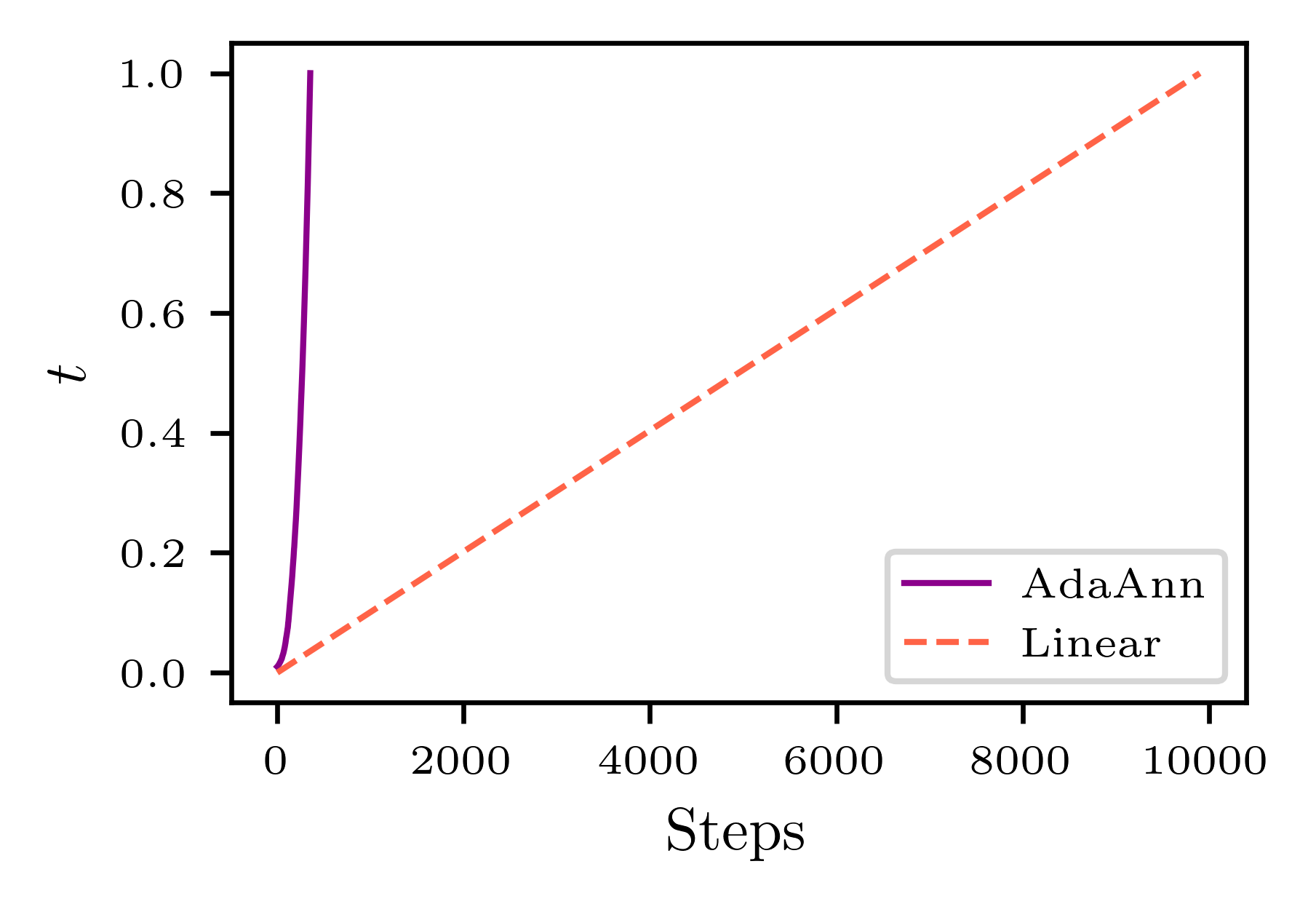} \\[-10pt]
\hbox{(a) AdaAnn} & \hbox{(b) AdaAnn and Linear}
\end{array}$
\caption{(a) AdaAnn annealing schedule and 
(b) comparison between the AdaAnn and a linear schedule for density approximation in Example 1.}
\label{fig:bimodal_schedules}
\end{figure}

The final optimized variational distribution via the planar flow with AdaAnn is presented in Figure~\ref{fig:bimodal_adaAnn}, which shows an accurate approximation of the target distribution.
Though both AdaAnn and the linear scheduler (Figure \ref{fig:bimodal_1mode}) perform well in approximating the target distribution in this example, the computational cost associated with the  linear scheduler is much higher.   The linear  schedule performed 9,902 steps with a total of 18,400 parameter updates in 12.50 minutes whereas AdaAnn required 354 steps with the total number of 9,204 parameter updates in 6.51 minutes as summarized in Figure~\ref{fig:bimodal_schedules}. The rate of change in $t_k$ in AdaAnn is slow when $t_k$ is small and increases with $t_k$; this adaptive behavior helps drive the computational cost down for AdaAnn.  These computations were performed on a laptop using a 1.80 GHz Intel Core i7-10510U processor.

\begin{figure}[!htb]
    \centering
$\begin{array}{cc}
    \includegraphics{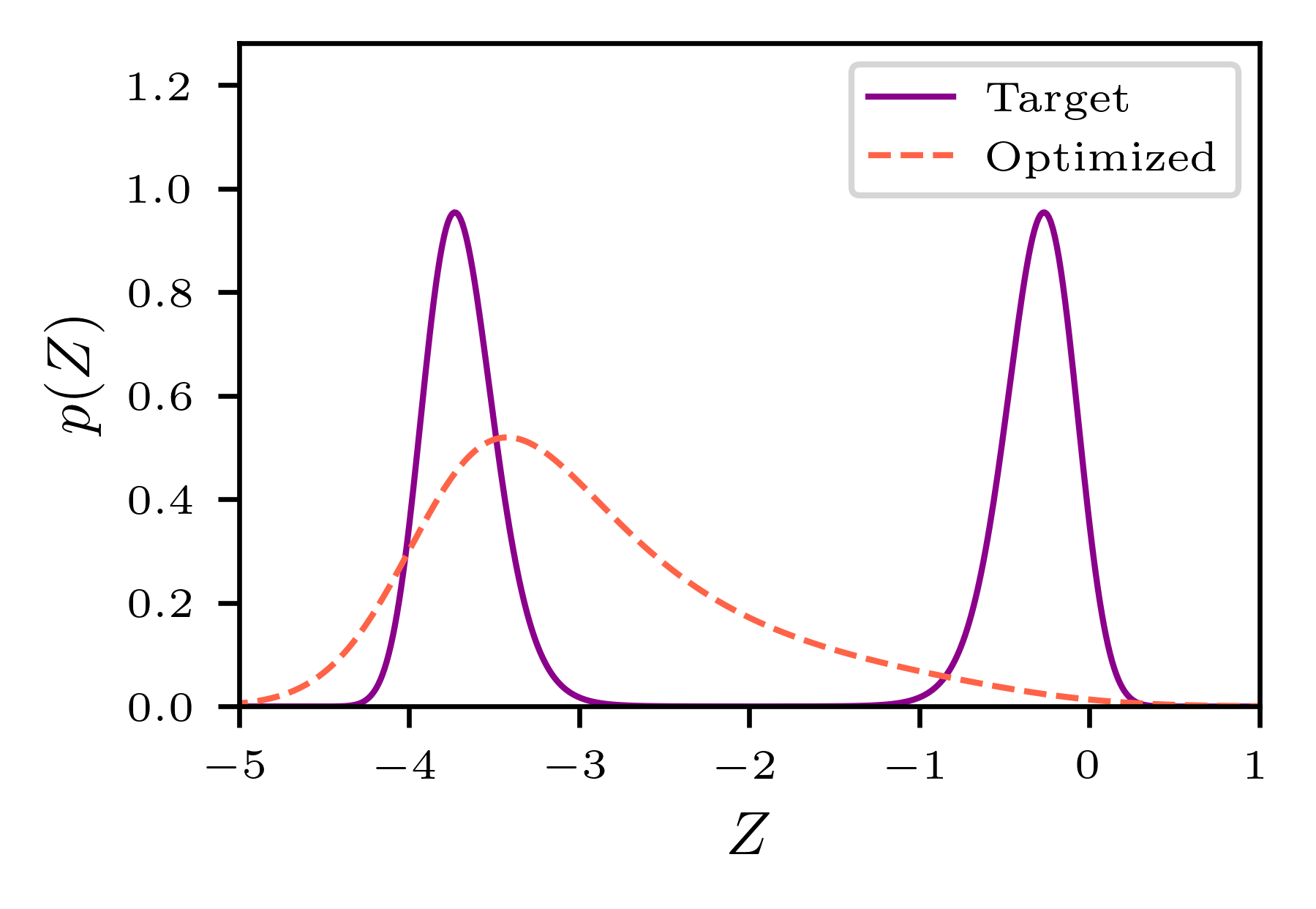}  &
    \includegraphics{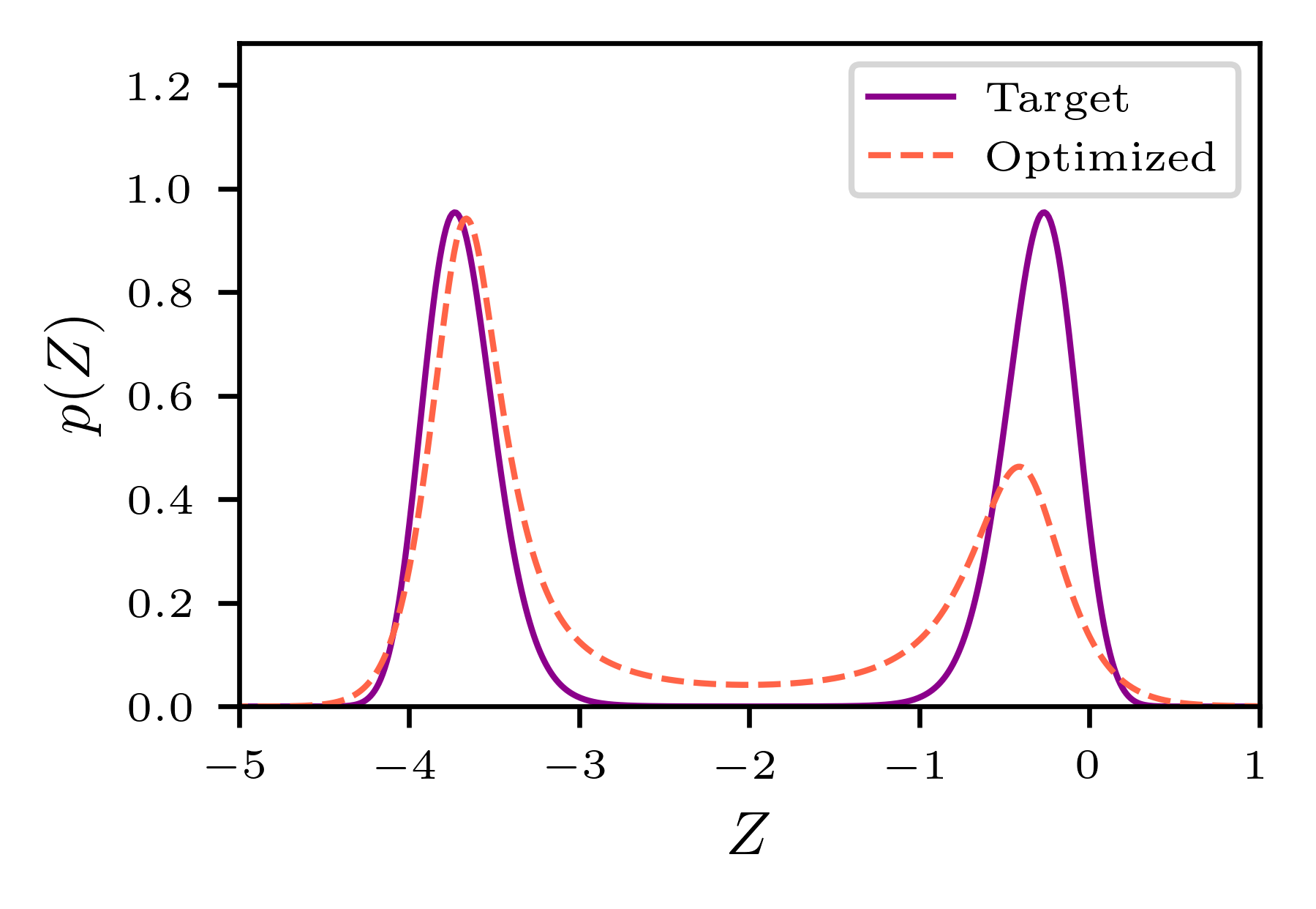} \\[-10pt]
    \hbox{(a) $\tau= 1$} & \hbox{(b) $\tau= 0.1$}\\
    \includegraphics{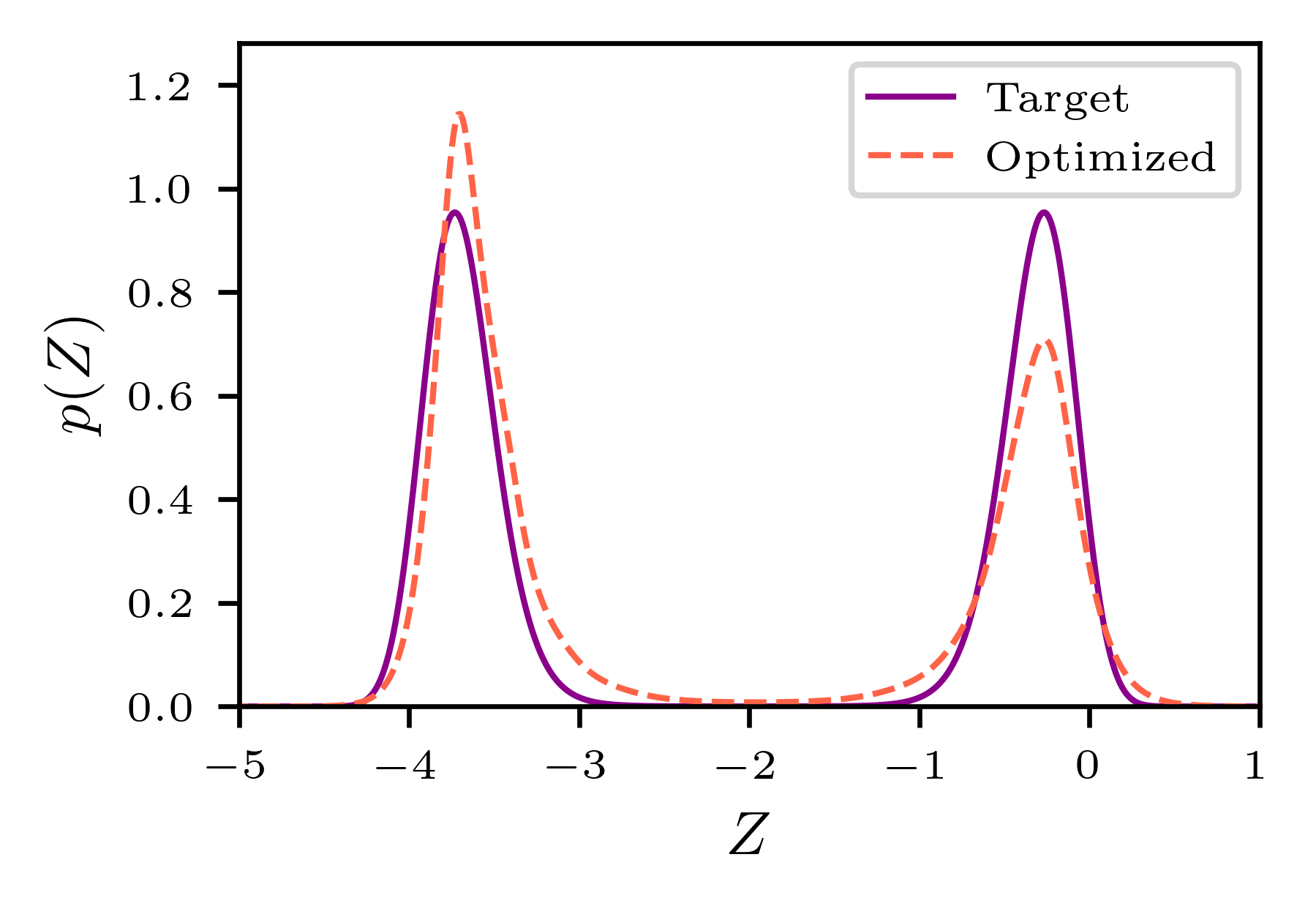} &
    \includegraphics{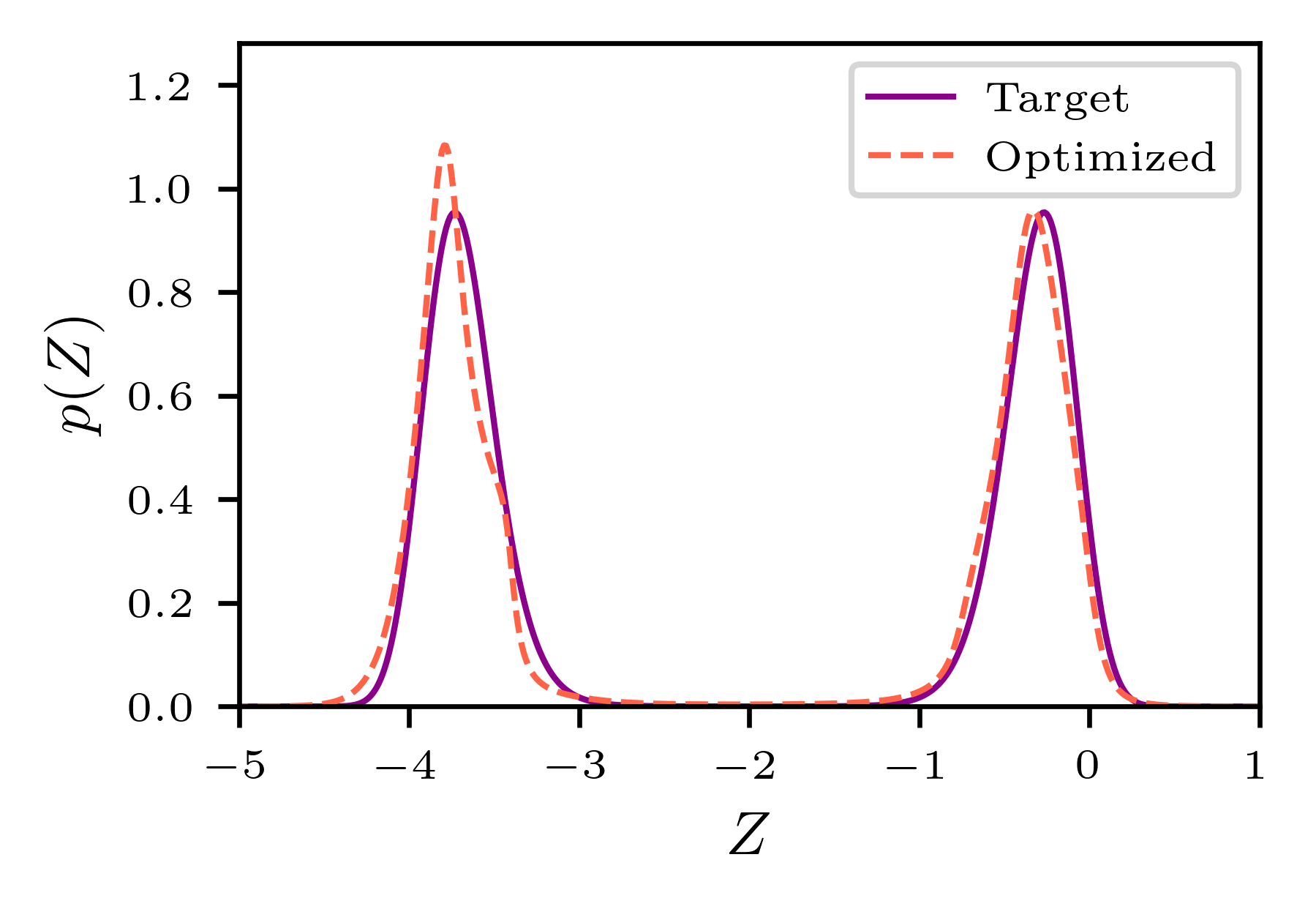} \\[-10pt]
    \hbox{(c) $\tau= 0.01$} & \hbox{(d) $\tau = 0.001$}
    \end{array}$
    \caption{Comparison of optimized distributions from NFs with AdaAnn at $t=1$ for various KL tolerances $\tau$.} \vspace{-6pt}
    \label{fig:choosing_tol}
\end{figure}

We also examine how the choice of $\tau$ in AdaAnn affects the approximation quality and computational complexity. Toward that end, we set the KL divergence tolerance $\tau$ at 4 different values (1, 0.1, 0.01, 0.001) and obtained the final approximate distribution to the target in each case. The results are presented in Figure~\ref{fig:choosing_tol}.
For a tolerance as large as $\tau=1$, the optimized density at $t=1$, after taking only 10 incremental temperature steps, is evidently biased toward one of the modes.
Continuing to train from this point, the approximation converges to a single mode. 
The next three smaller tolerance choices manage to maintain the bimodal structure. For $\tau=0.1$, the annealing phase completes in 56 temperature incremental steps but still needs a decent number of refinement steps to provide a good approximation for the target density; $\tau=0.01$ takes 355 steps; and $\tau=0.001$ takes 2,929 steps in the annealing phase. While $\tau=0.001$ provides a slightly better approximation than $\tau=0.01$, it takes significantly more steps (8.25 folds more) without significantly improving the quality of the resulting approximation.

In summary, this example illustrates that while AdaAnn and the linear annealing schedule lead to favorable approximations to the target distribution, AdaAnn significantly reduces the number of steps needed to the final approximation and ultimately reduces the computational time. In addition, the choice of the KL divergence tolerance $\tau$ is critical for the accuracy of the variational approximation: too large a $\tau$ value can be too crude to capture important characteristics of a distribution (e.g., multi-modality) and too small a $\tau$ value may incur additional computational costs without significantly improving the approximation.

\subsection{Example 2: One-dimensional Mixture Gaussian Distribution}\label{sec:1D_Case_Study}
\begin{figure}[!htb]
    \centering
    \includegraphics{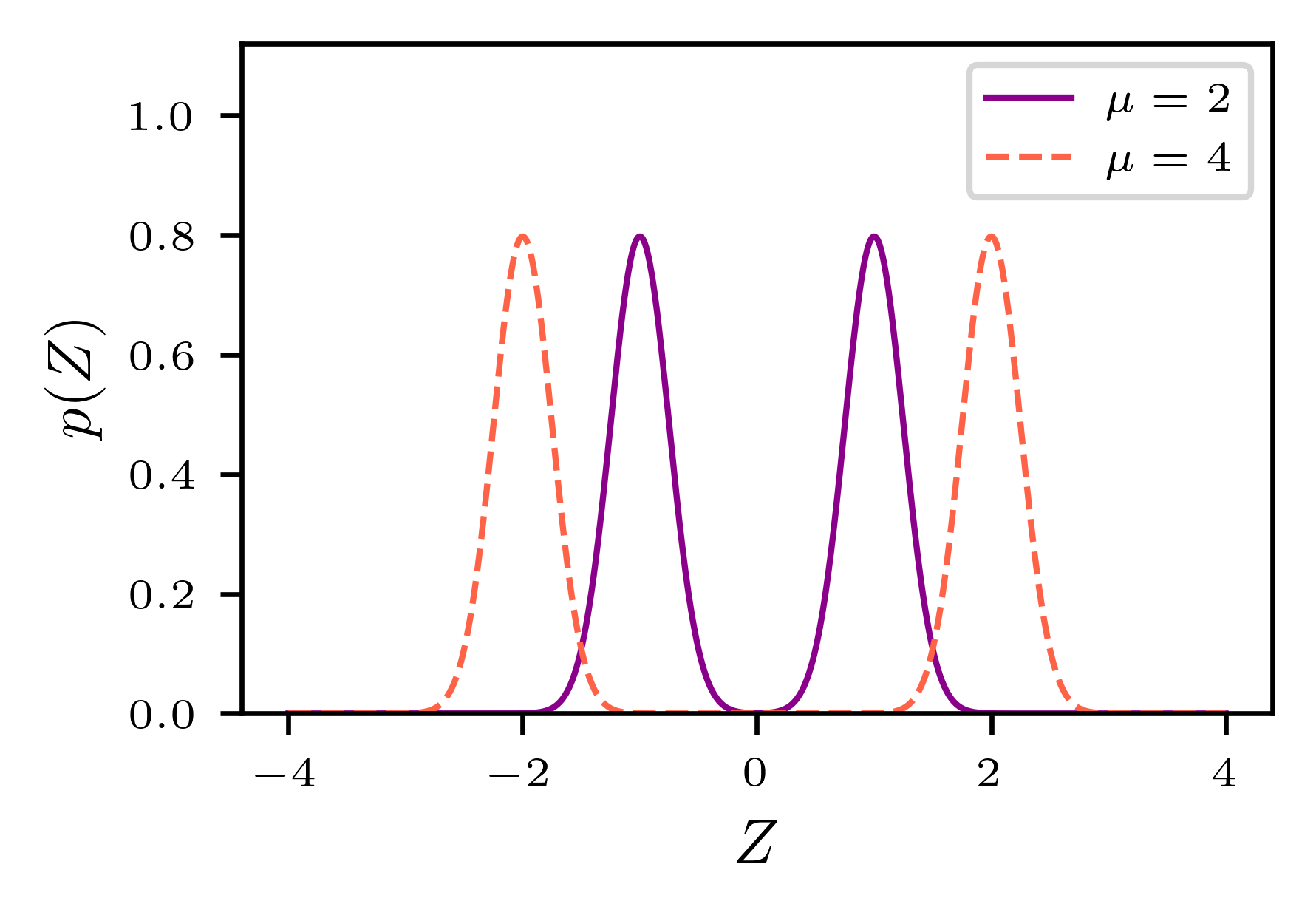} 
    \includegraphics{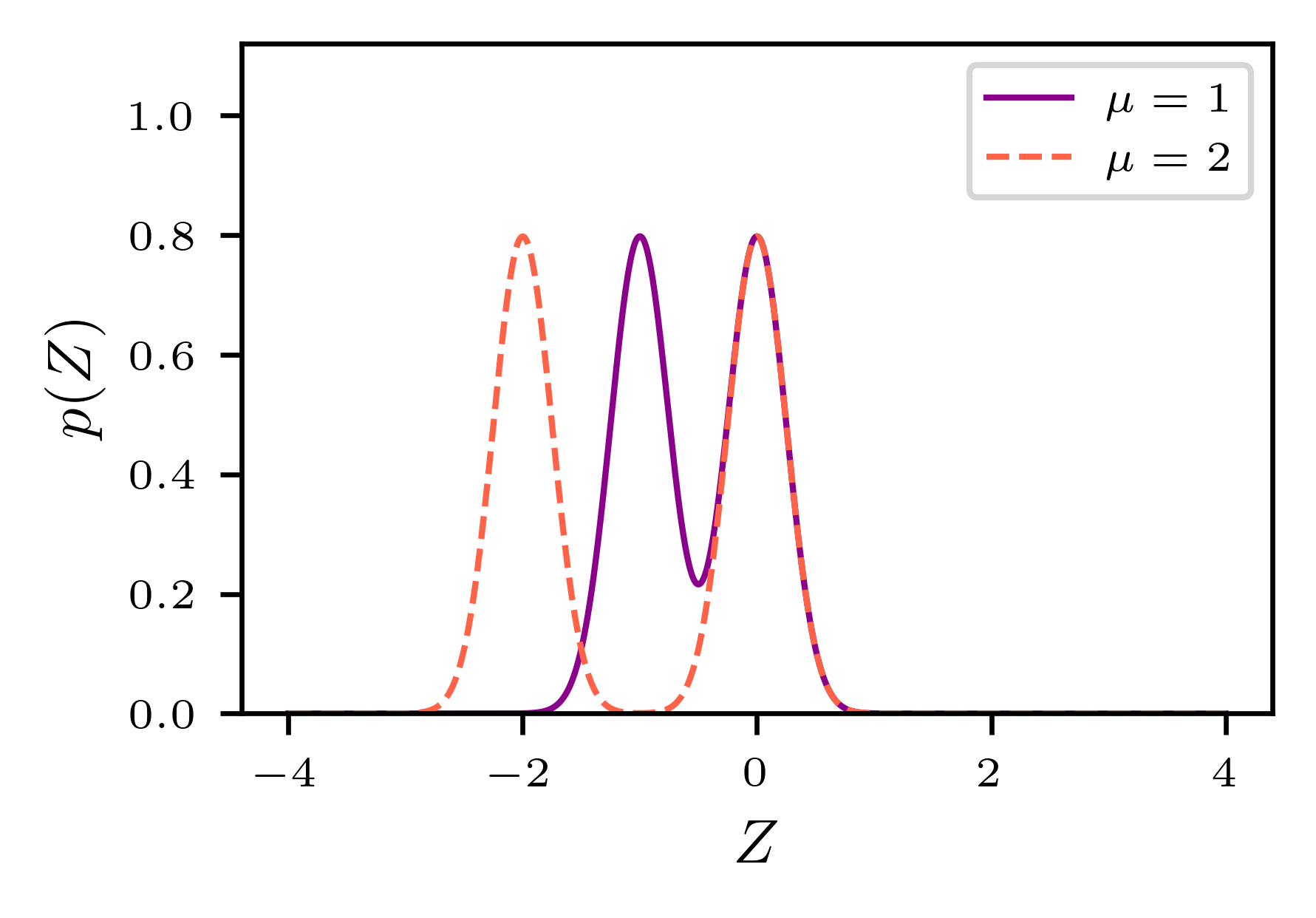}\vspace{-12pt}
    \caption{Illustration of symmetric and asymmetric bimodal distributions in example 2.}\vspace{-6pt}
    \label{fig:GMMCases}
\end{figure}

We consider a mixture of two Gaussian distributions in this example, namely
\begin{equation}\label{eq:GMM}
    p(Z) = \frac{1}{2 \sqrt{\pi/8}} e^{-8(Z + \mu_1 )^2} + \frac{1}{2 \sqrt{\pi/8}} e^{-8(Z + \mu_2 )^2}.
\end{equation}
Here, $p(Z)$ depends upon two parameters, $\mu_1$ and $\mu_2$, which are varied to investigate how the distance between the two modes of $p(Z)$ and their location relative to the mode of the base distribution $q_{0}=\mathcal{N}(\mu=0,\sigma^2=16)$ impact the accuracy of the optimal variational approximation.
We examine two cases of $p(Z)$: (1) when the two modes are symmetrically located around 0, the mode of $q_0$; that is, $\mu_1 = -\mu_2$, and (2) when one of the modes is fixed at 0. 
We refer to these two cases as the \emph{symmetric} and \emph{asymmetric} case, respectively, and use a single parameter $\mu$ to denote the distance between the two modes in both cases.
For the symmetric case, we set $\mu_1=\mu/2$ and $\mu_2=-\mu/2$ while, for the asymmetric case, $\mu_1=\mu$ and $\mu_2=0$. 
An example for each of the two cases is provided in Figure~\ref{fig:GMMCases}.

We vary $\mu$ from 1 to 16 in the symmetric case and 1 to 8 in the asymmetric case. For each value of $\mu$, we run 50 trials without annealing, with a linear annealing schedule, and with the AdaAnn scheduler to approximate the target distribution for VI via NFs. The number of layers for the planar flow is $L=50$  for the symmetric and $L=75$ for the asymmetric case. 
For the case without annealing, we also vary the number of layers and consider $L=25,50,75$ and train the planar flow for 8,000 iterations with $N=100$ samples per iterations. We set $t_0=0.01$, $\tau=0.005$, $M=1,000$, $T_0=500$, $T=5$, and $N=100$ for AdaAnn (Algorithm \ref{alg:adaptive})
with $\epsilon=10^{-4}$, $T=1$, $N=100$ for the linear scheduler.
For the purposes of this example, we do not be further refining the solution at $t=1$, but rather indicating whether the optimized distribution contains a bimodal structure.   The learning rates for the Adam optimizer are reported in Table~\ref{table:1D_caseStudy_learningRates}.
\begin{table}[!htb]
\caption{Learning rate for the Adam optimizer for different values of $\mu$ in example 2.}
\centering
    \begin{tabular}[2in]{c | c|c|ccc| ccc ccc| c|c|c|c cccc}
    \toprule
    &\multicolumn{11}{c|}{symmetric case ($\mu_1=\mu/2, \mu_2=-\mu/2$)} &
    \multicolumn{8}{c}{asymmetric case ($\mu_1=\mu, \mu_2=0$)}\\
    \hline
    $\mu$ & 1 & 2 & 3 & 4 & 5 & 6 & 8 & 10 & 12 & 14 & 16 & 1 & 2 & 3 & 4 & 5 & 6 & 7 & 8\\
     \hline
    & 0.02 & 0.02 & \multicolumn{3}{c|}{0.001} & \multicolumn{6}{c|}{0.0005}
    & 0.01 & 0.01 & 0.002 & \multicolumn{5}{c}{0.001}\\
    \bottomrule
    \end{tabular}
    \label{table:1D_caseStudy_learningRates}
\end{table}

\begin{figure}[!htb]
    \centering
    $\begin{array}{cc}
    \includegraphics[scale=0.97]{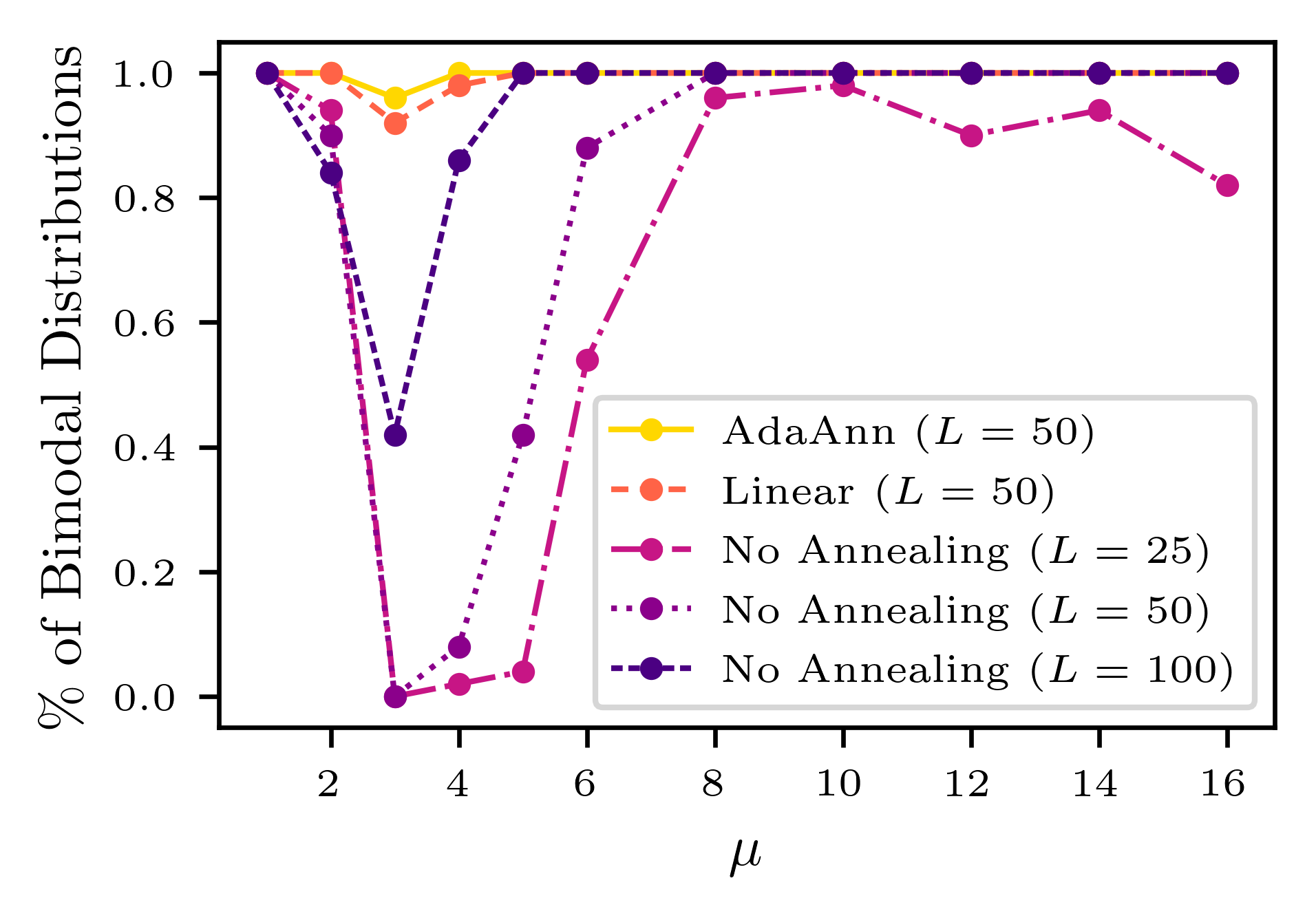} &
    \includegraphics[scale=0.97]{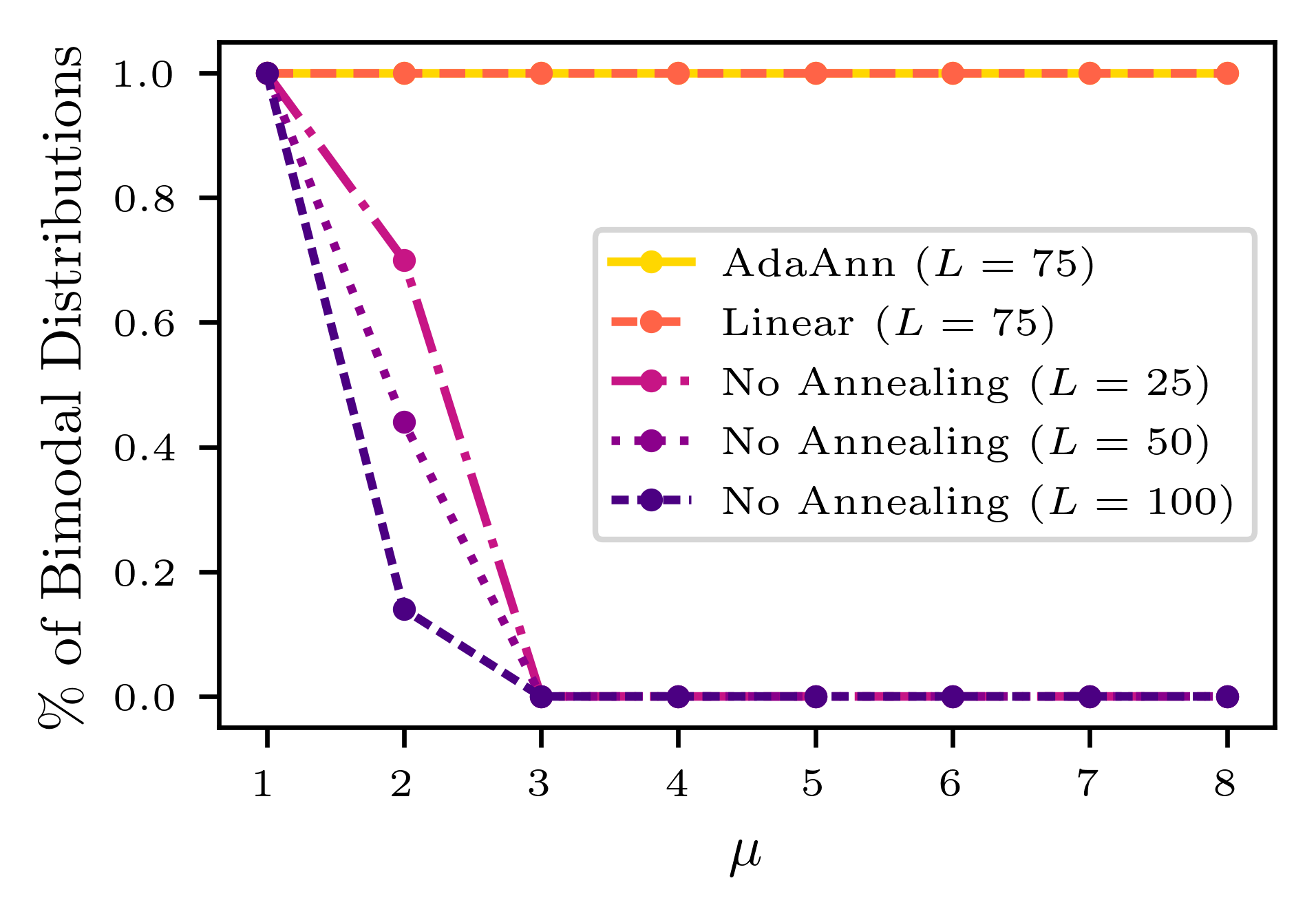} \\[-10pt]
    \hbox{(a) Symmetric} & \hbox{(b) Asymmetric}
    \end{array}$
    \caption{Percentage of approximate distributions from VI via NFs which capture the bimodal target structure out of 50 trials in Example 2.}
    \label{fig:GMMresults}
\end{figure}

We examine the number of repetitions where the approximated distributions of VI via NFs that 
captures the bimodal structure in $p(Z)$. The results are summarized  in Figure~\ref{fig:GMMresults}.  We also compared the computational time required by NFs without annealing, by the proposed AdaAnn scheduler, and by a linear scheduler (the computations were performed on a Intel(R) Xeon(R) CPU E5-2680 v3 @ 2.50GHz Haswell processors with 256 GB of RAM).  The results in the symmetric case at $\mu=2$ are presented in more detail in Table~\ref{table:1D_caseStudy_time_Annealing}. 

First, for the symmetric case with $\mu=4$, an increasing number of planar flow layers is associated with a higher likelihood to capture the target bimodal structure without annealing, but at higher computational cost (Table~\ref{table:1D_caseStudy_time_Annealing}). 
Second, without annealing, the symmetric modes are well recovered for $\mu>1$; for the asymmetric case, the two modes are captured 100\% of the time only when $\mu = 1$, independent on the number of layers, and the percentage decreases with $\mu$, until reaches 0 for $\mu\ge3$.
Third, with annealing, the target distributions can be accurately approximated for all the examined values of $\mu$ in both the symmetric and asymmetric cases.
Considering the symmetric case without annealing, there is a large drop in the percentage of recovered bimodal distributions near $\mu=4$ for $L=25$ and 50. When the modes of the target distribution are \emph{connected}, i.e., not separated by a segment of zero probability, NFs easily captures both modes. 
When the modes become separated, NFs no longer capture both modes consistently. This is indicative of a rough loss landscape where the optimizer is unable to determine the global minimum. As these modes become further separated, NFs improve in capturing both modes, likely indicating the loss landscape has become smoother and the global minimum is easier to attain.
Though both the linear and AdaAnn annealing schedules are able to produce bimodal approximations of similar accuracy, AdaAnn requires significantly fewer parameter updates than a linear scheduler  (Table~\ref{table:1D_caseStudy_time_Annealing}).
\begin{table}[!htb]
\caption{Computational time (in minutes) required for VI-NF with and without annealing, applied to the symmetric case at $\mu=4$ in Example 2.}
\centering
    \begin{tabular}[2in]{lccc}
    \toprule
    \bf{NF procedure} & \bf{5th-Percentile} & \bf{Median} & \bf{95th-Percentile} \\
    \midrule
    No annealing ($L=25$) & 2.80 & 2.84 & 2.97 \\
    No annealing ($L=50$)& 5.63 & 5.71 & 6.24 \\
    No annealing ($L=75$) & 10.87 & 11.28 & 11.98 \\
    Linear schedule ($L=50$) & 6.45 & 7.23 & 7.75 \\
    AdaAnn ($L=50$) & 3.04 & 3.25 & 3.81 \\
    \bottomrule
    \end{tabular}
    \label{table:1D_caseStudy_time_Annealing}
\end{table}

In summary, the results suggest that, for NFs without annealing, (1) the relative location of the base distribution with respect to the locations of the modes of the target distribution may affect the accuracy of the variational approximation and (2) when the location of the base distribution is strongly biased toward one of the modes of the target distribution, successful approximation may only occur when the modes are not separated ($\mu\le1$ in this example). Annealing helps to mitigate both problems. 

\subsection{Example 3: Two-dimensional  Bimodal Distribution}

\begin{figure}[!htb]
    \centering
    $\begin{array}{ccc}
    \includegraphics[height=1.9in]{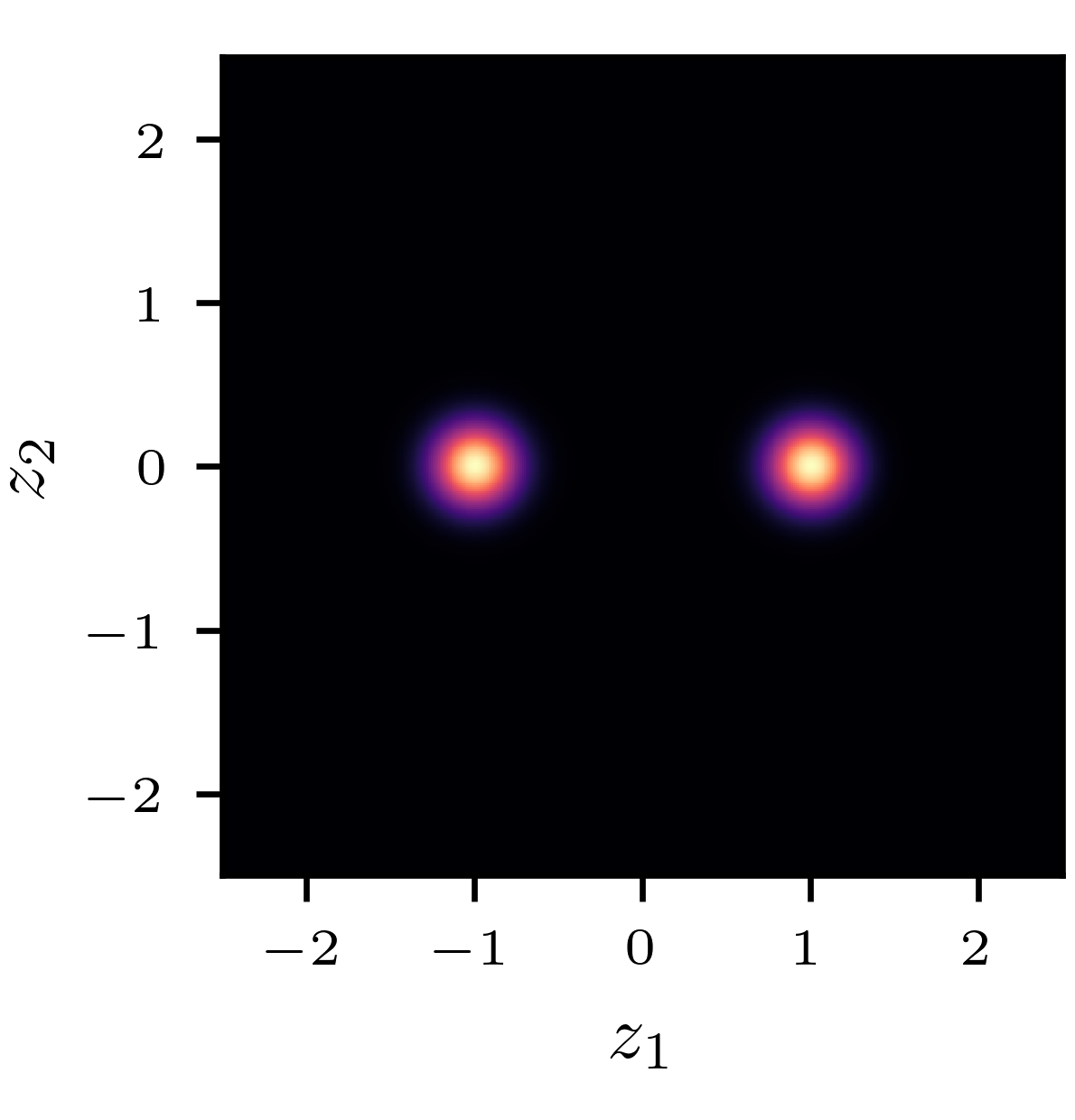} & 
    \includegraphics[height=1.9in]{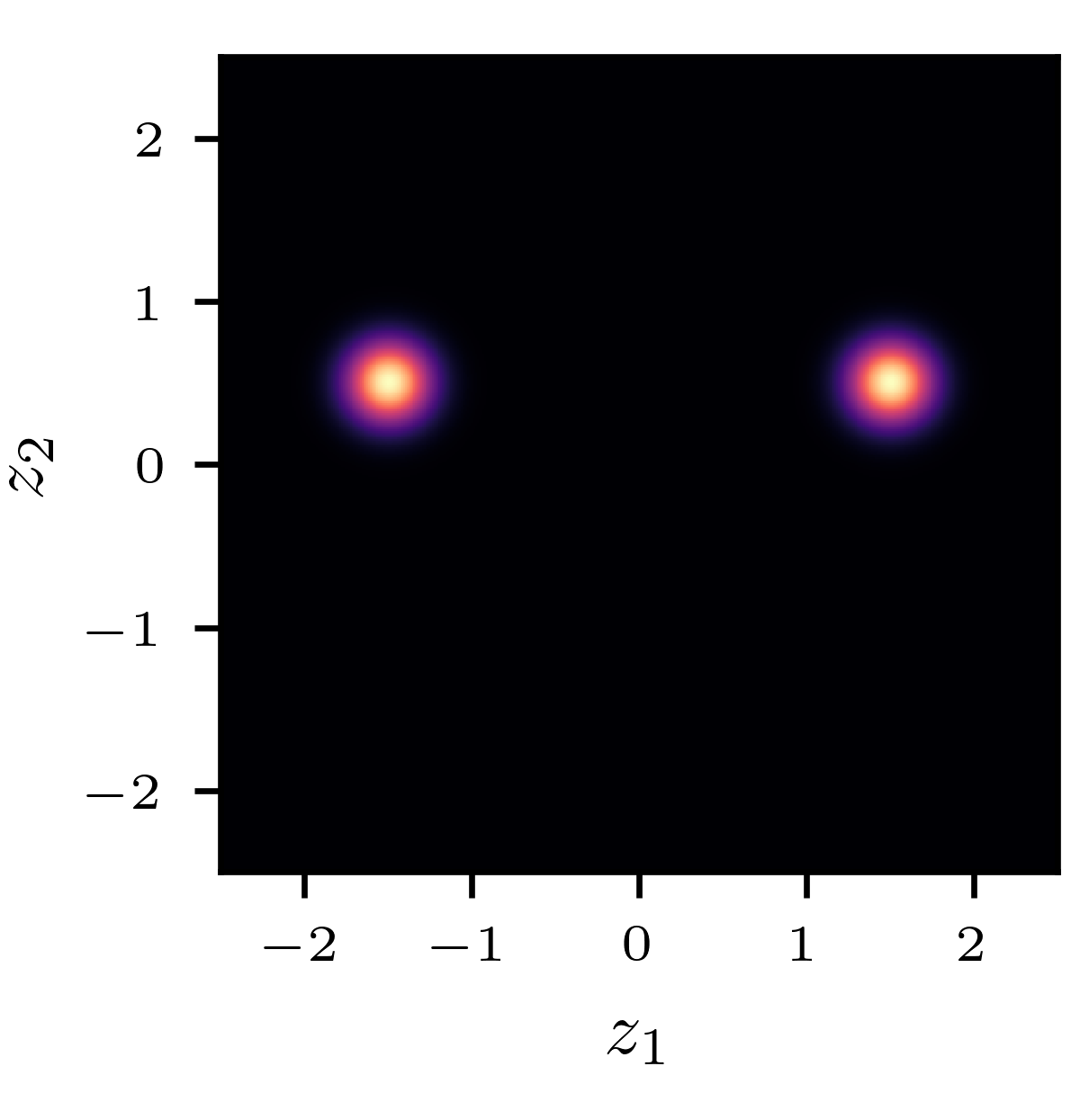} & \includegraphics[height=1.9in]{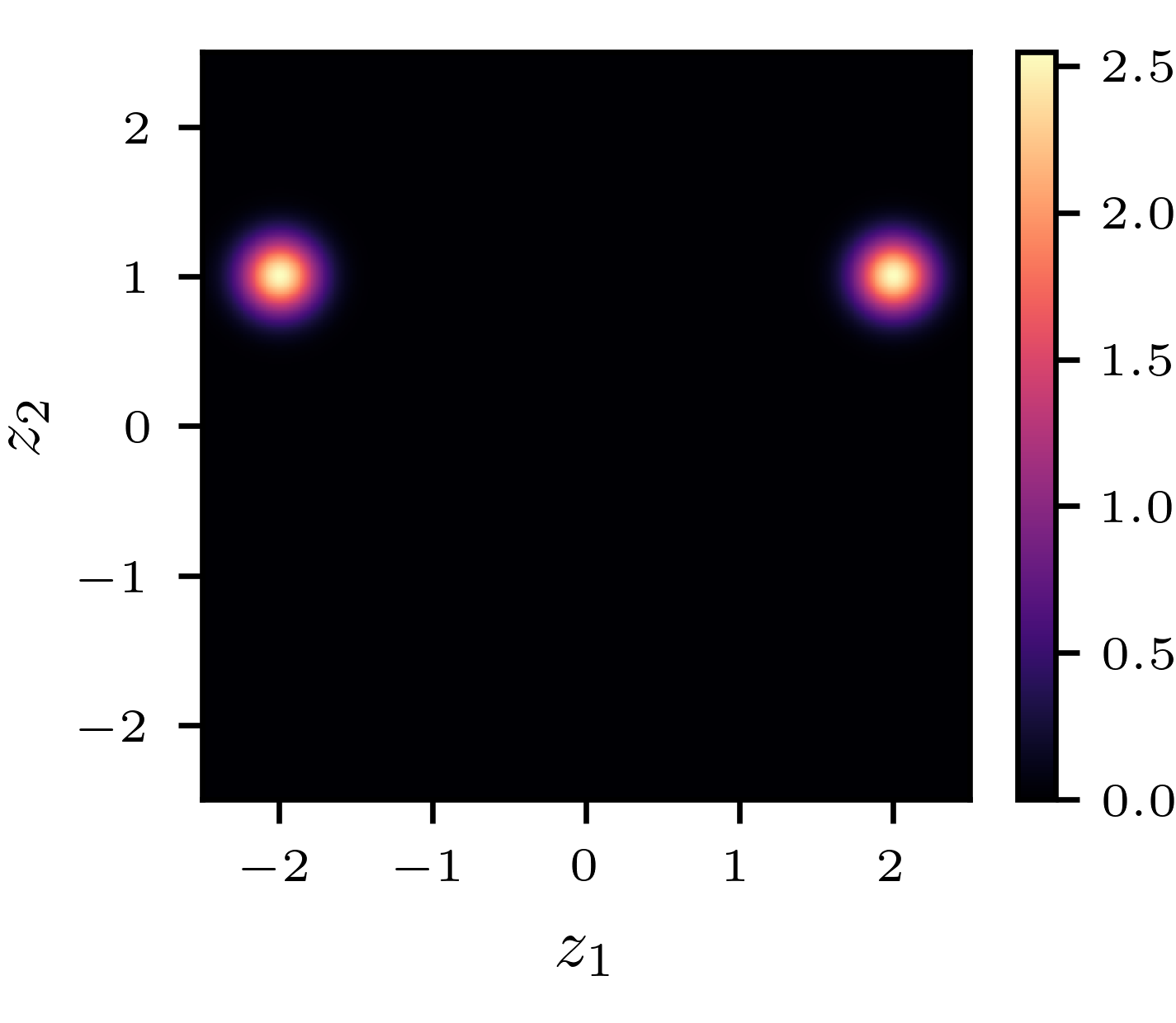} \\[-10pt]
    \hbox{(a) $\mu=0$} & \hbox{(b) $\mu=0.5$} & \hbox{(c) $\mu=1$}
    \end{array}$
    \caption{Bivariate Gaussian mixture densities for increasing values of $\mu$ in Example 3.}
    \label{fig:2D_caseStudy_distribution}
\end{figure}

In the third example, we compare the density approximation performance between planar flows coupled with AdaAnn and a more expressive flow such as realNVP. 
The target distribution is a mixture of two bivariate Gaussian densities expressed as
\begin{equation}\label{eq:GMM2D}
    p(Z_1, Z_2) = \frac{8}{\pi} e^{-16\,[(Z_1 + \mu + 1)^2 + (Z_2 - \mu)^2]} + \frac{8}{\pi} e^{-16\,[(Z_1 - \mu - 1)^2 + (Z_2 - \mu)^2]}.
\end{equation}
Here, $p(Z_1,Z_2)$ depends upon a parameter $\mu$
which is used to move the modes.
In particular, this density is similar to the bimodal symmetric density from Section~\ref{sec:1D_Case_Study} and has narrow modes equally spaced from the origin. 
As $\mu$ increases, the modes will move diagonally up and away from the origin resulting in a larger separation, as seen in Figure~\ref{fig:2D_caseStudy_distribution}.

To approximate the target distribution $p(Z_1, Z_2)$, we transform $q_{0}=\mathcal{N}(\mu=[0,0], \Sigma=4 I_2)$ using four different configuations: (1) three planar flows consisting of $L=50, 75, 100$ layers without annealing; (2) a planar flow with $L=75$ layers combined with the AdaAnn scheduler ($\tau=0.01, M=1,000, t_0=0.01, T_0=500, T=5$, no refinement at $t=1$); (3) a planar flow with a linear scheduler ($t_0=0.01, \epsilon=10^{-4}, T_0=500, T=1$, no refinement at $t=1$); and (4) realNVP without annealing. 
For realNVP, the scale and translation function $\bm{a}_s$ and $\bm{a}_t$, respectively, 
consist of fully connected neural networks with two neurons for both the inputs and output layers, two hidden layers with $H$ hidden neurons, and the ReLU activation function. A hyperbolic tangent activation function is applied right before the output layer on the scale function $\bm{a}_s$. We examined three cases 
for $H$, namely 10, 25, and 100.
We examined two scenarios of coupling layers,
namely 6 and 12, and use alternating masking that switches the variables being updated at each coupling layer.  
We trained the realNVP for 5,000 iterations. For the Adam optimizer, we used a batch size of $N=100$ and the learning rates are shown~in~Table~\ref{table:2D_caseStudy_learningRates}. 
\begin{table}[!htb]
\caption{Learning rates for different values of $\mu$ used for Example 3.}
\label{table:2D_caseStudy_learningRates}
\centering
    \begin{tabular}[2in]{c | ccccccc}
    \toprule
    $\mu$ & 0.0 & 0.5 & 1.0 & 1.5 & 2.0 & 2.5\\
    \midrule
    learning rate for adam optimizer & 0.001 & 0.0008 & 0.0005 & 0.0005 & 0.0005 & 0.0002\\
    \bottomrule
    \end{tabular}
\end{table}

\begin{figure}[!htb]
    \centering
    $\begin{array}{cc}
    \includegraphics{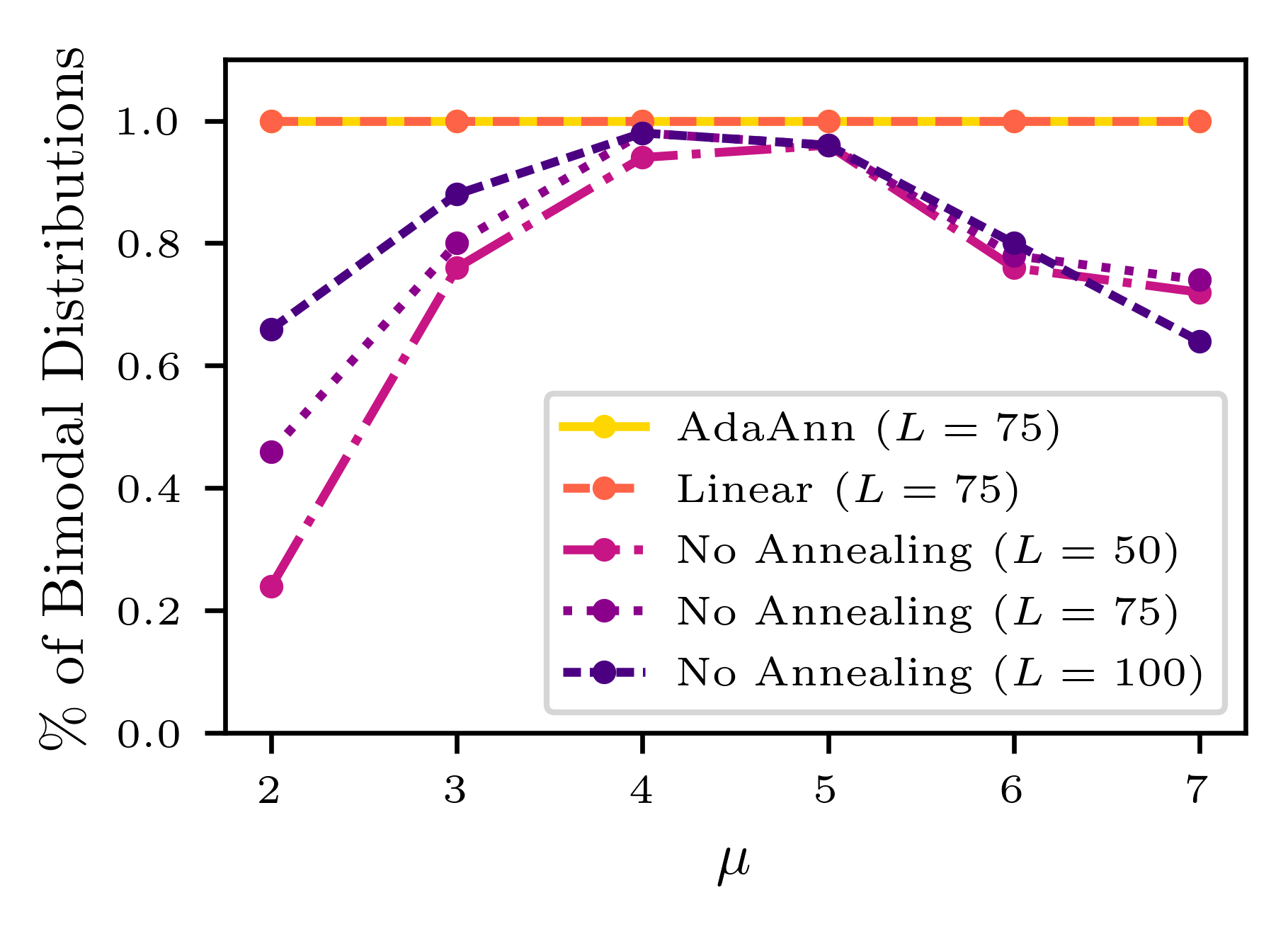} & 
    \includegraphics{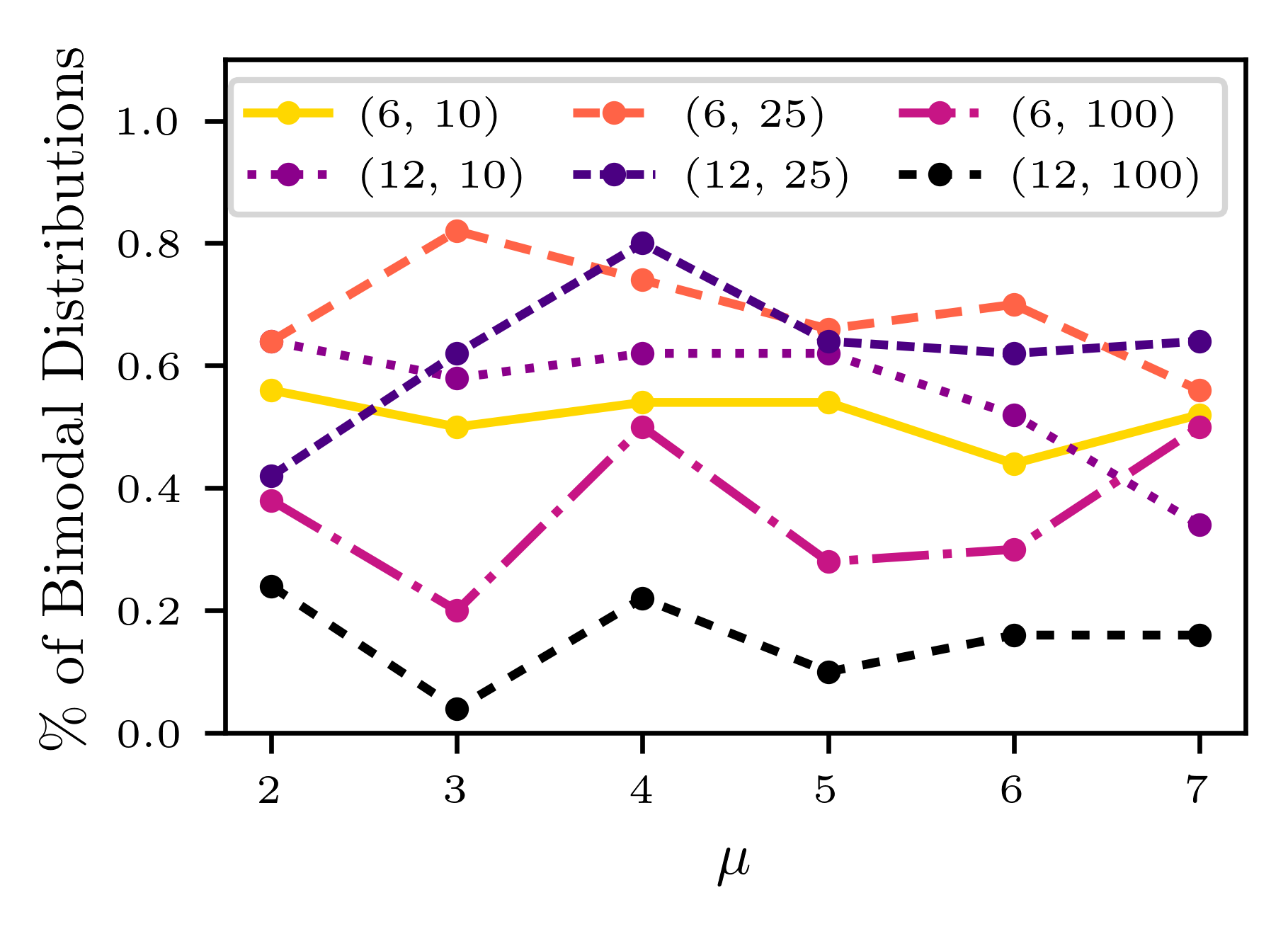} \\[-10pt]
    \hbox{(a) Planar Flow} & \hbox{(b) RealNVP (\# of layers, \# of neurons)}
    \end{array}$
    \caption{Rate of successful distribution reconstruction in Example 3.}
    \label{fig:2D_caseStudy_results}
\end{figure}

For each $\mu$ and each NF setup, we conducted 50 trials and recorded how many times the bimodal structure of the target distribution is captured in the final optimized distribution. 
The results are summarized in Figure~\ref{fig:2D_caseStudy_results}.
In particular, both of the annealing methods capture the bimodal structure in all 50 trials at every $\mu$, outperforming the planar flows without annealing,
which is consistent with the results from
Examples 1 and 2.
RealNVP, despite having a more complicated structure than planar flow, still fails to capture both modes in a considerable number of repetitions, suggesting that the approximation accuracy resulting from planar flow plus an annealing schedule may not be achieved by a more expressive~flow~alone.

Both annealing methods with planar flows achieve the same accuracy, but AdaAnn has significantly fewer parameter updates during the annealing phase
yielding superior computational efficiency. For example, at $\mu=0.5$, the median number of parameter updates is 2,575, with the $5^{\rm th}$ and $95^{\rm th}$ percentiles being 2,485 and 2,748 parameter updates, respectively, for AdaAnn compared to 10,401 for the linear schedule.

\subsection{Example 4: Lorenz Attractor}

After considering closed-form distributions in 
the first three examples, we 
investigate the ability of VI via NFs 
with annealing to solve inverse problems involving dynamical systems. 
In such cases, evaluating the posterior distribution at a single realization of the input parameters (up to a constant) necessitates the numerical solution of a system of ordinary differential equations (ODEs).
Specifically, in this section we consider the Lorenz attractor~\cite{DeterministicNonperiodicFlow}:
\begin{equation}\label{eq:Lorenz}
    \begin{cases}
        \dot{x} = s(y-x)\\
        \dot{y} = x(r-z) - y\\
        \dot{z} = xy - bz.\\
    \end{cases}
\end{equation}
This system of ODEs results from a simplified representation of Rayleigh-B\'enard convection and is derived from a Galerkin projection of a system of coupled Navier-Stokes and heat transfer equations with thermal convection and buoyancy. 
It models convection between two horizontal plates with the lower plate uniformly warmer than the upper plate. Described by this system, $x$ is proportional to the intensity of the convective motion, $y$ is proportional to the temperature difference between ascending and descending currents, and $z$ is proportional to the discrepancy between the vertical temperature distribution in the model and a linear profile \cite{DeterministicNonperiodicFlow}. Restricted to positive values, $s$ is the Prandtl number, $r$ is the Rayleigh number, and $b$ is a geometric factor, i.e., the aspect ratio of the convection vortices \cite{DeterministicNonperiodicFlow, strogatz2000nonlinear}. The system is unstable for $\sigma > (b+1)$ and $r > r_{c}\approx 24.74$.
In particular, for $s = 10$, $b = 8/3$, and $r = 28$, it follows a chaotic butterfly-like dynamics revolving around two strange attractors. Starting from almost identical initial conditions, the system is known to generate chaotic trajectories for $t>15$ \cite{vulpiani2009chaos}. 

\begin{figure}[!htb]
    \centering
    $\begin{array}{cc}
    \includegraphics[width=3in]{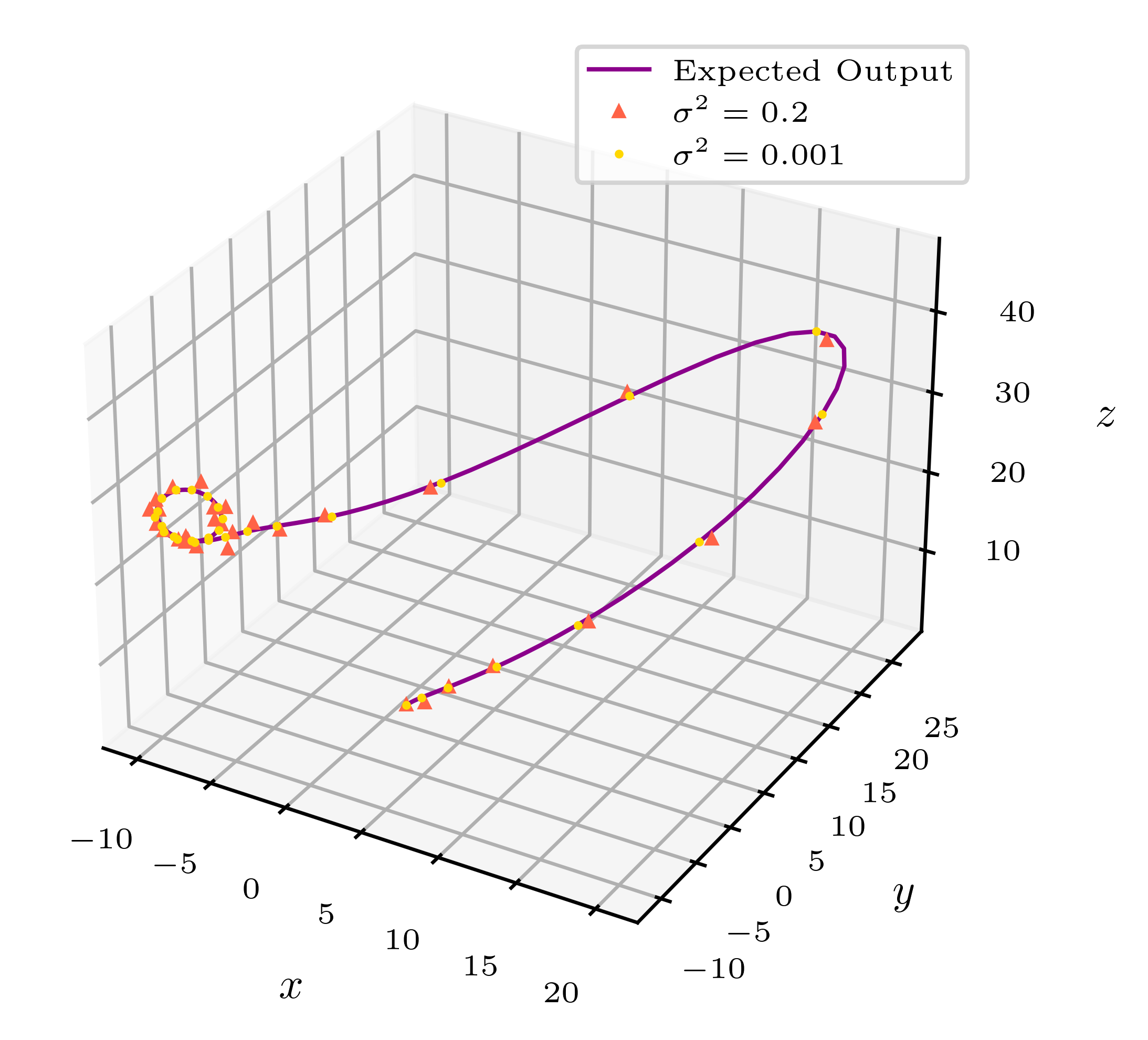} & 
    \includegraphics{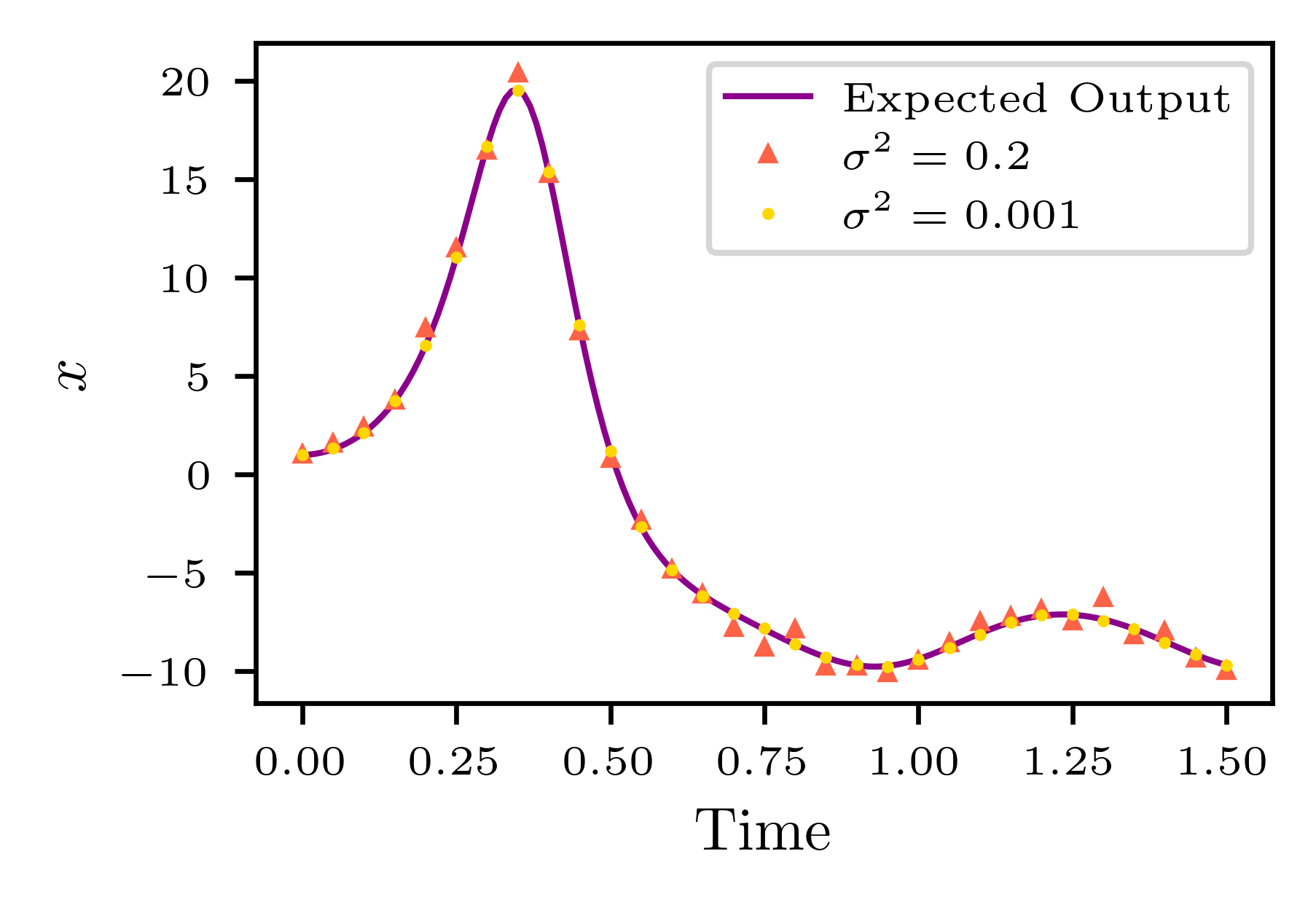} \\
    \hbox{(a) 3D Trajectory} & \hbox{(b) Trajectory of $x$ }\\    \includegraphics{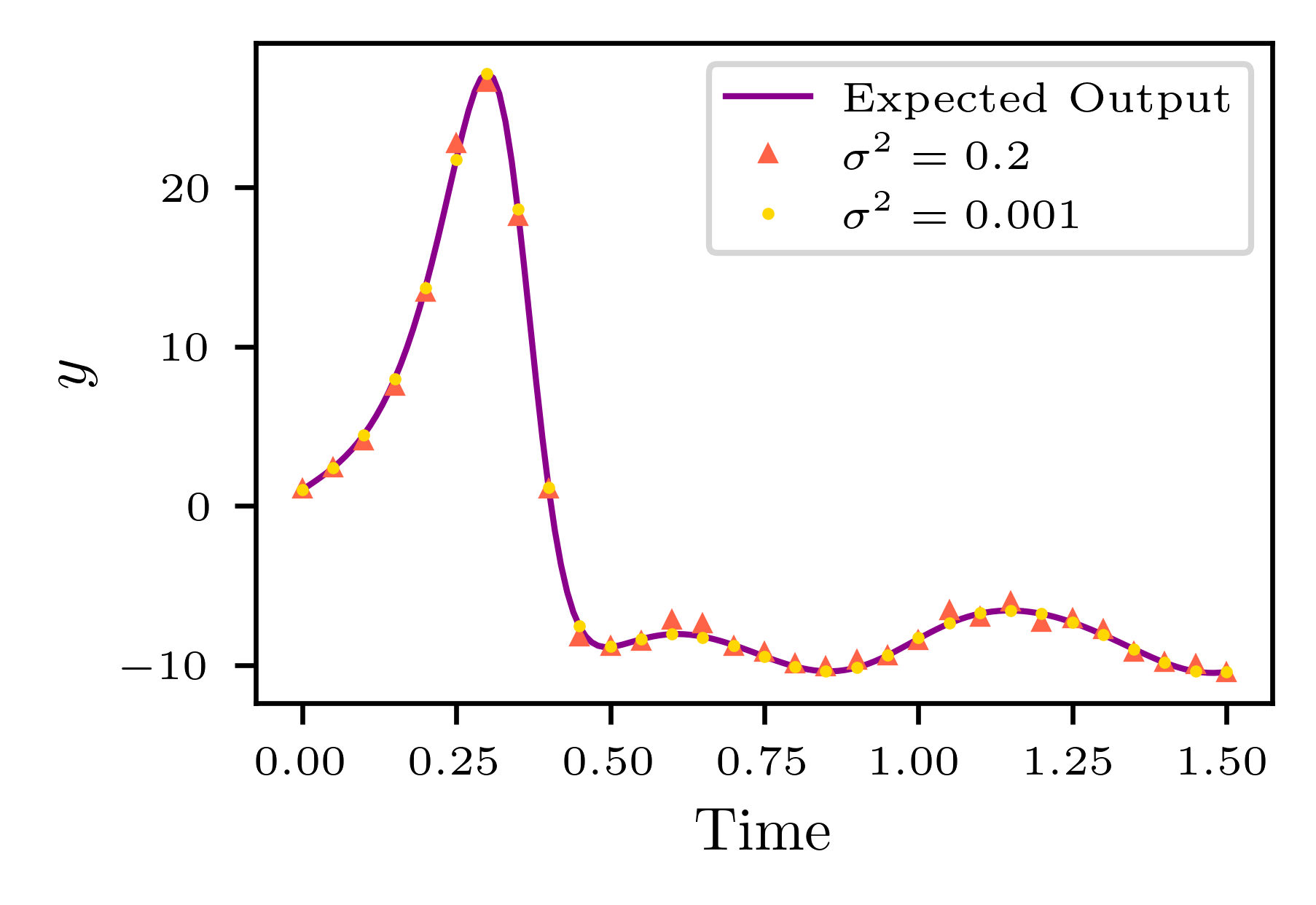} & 
    \includegraphics{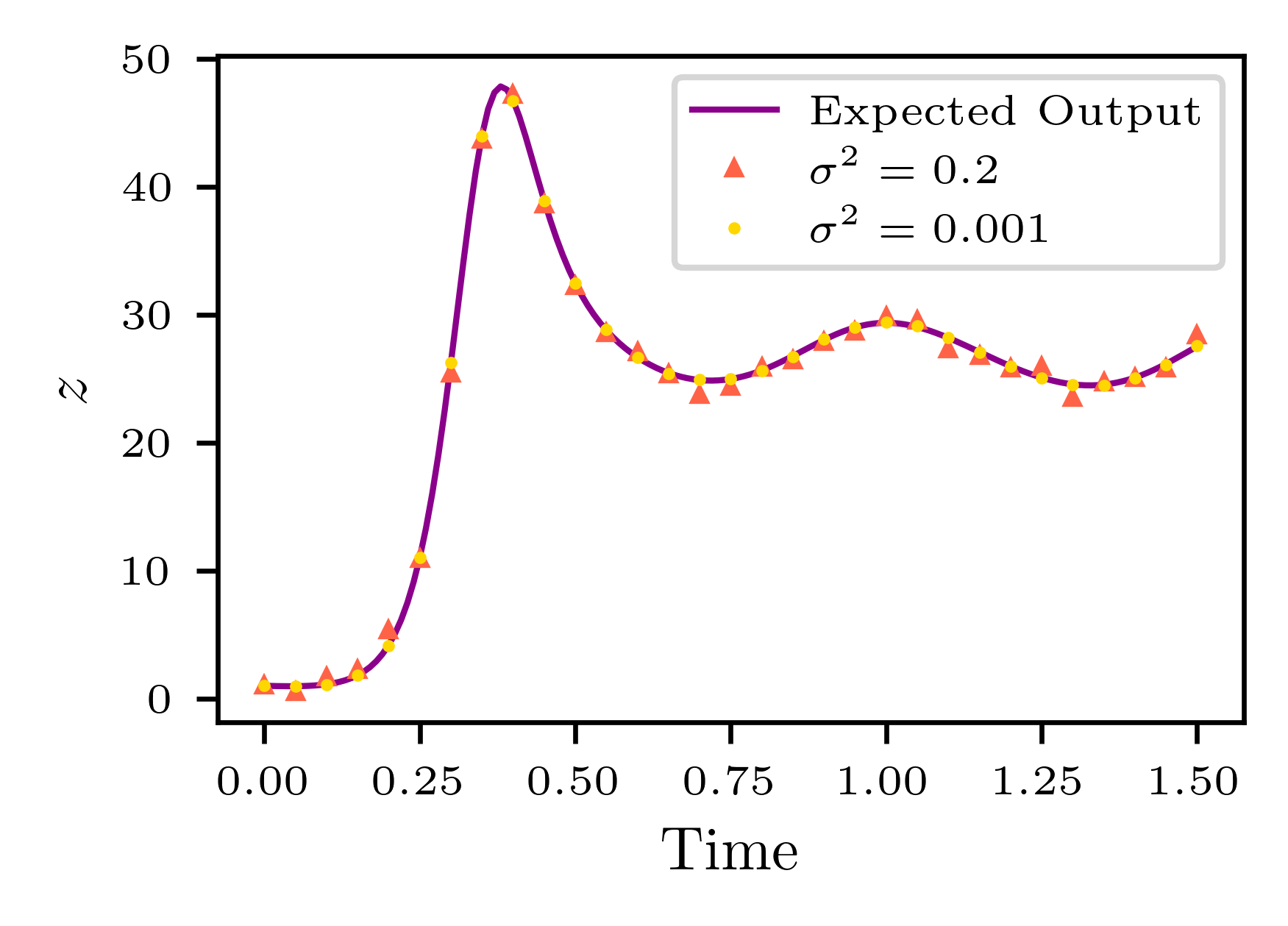} \\
    \hbox{(c) Trajectory of  $y$ } & \hbox{(d) Trajectory of $z$ }
    \end{array}$
    \caption{Trajectories of the Lorenz system and observations $(x,y,z)$.}
    \label{fig:Lorenz_trajectories}
\end{figure}

The parameters $s$, $b$, and $r$ in Eq.~\eqref{eq:Lorenz} are often of inferential interest given a set of observations on $x$, $y$, and $z$.  
We use VI via NFs to estimate  $s$, $b$, and $r$ 
in a Bayesian framework with a regime where the stable trajectories do not negatively affect the identifiability of the parameters. 
Specifically, we simulate observations given $s = 10$, $b = 8/3$, and $r = 28$ as follows. Using a fourth order Runge-Kutta method (RK4) with initial conditions
$x_0 = y_0 = z_0 = 1$, the Lorenz equations are integrated in time from $t=0$ to $t=1.5$ 
with step size $\Delta t=0.025$.  
From this solution $[(x_{i}, y_{i}, z_{i})]_{i=1}^{60}$, we choose $n = 30$ equally spaced data points and add Gaussian noise $\mathcal{N}(\mu = [0,0,0], \Sigma=\sigma^2 \cdot I_3)$ with $\sigma^2 = 0.001$ and $\sigma^2 = 0.2$, generating two sets of noisy $(x,y,z)$ realizations
as shown in Figure~\ref{fig:Lorenz_trajectories}.

The following is the posterior distribution of 
the parameters $\bm{\theta}=\{s, b, r\}$ 
with a non-informative uniform prior on the parameters and Gaussian likelihood function:
 \begin{equation}\label{eq:likelihood}
    p(\bm\theta|\bm (x,y,z)) \propto \dfrac{1}{\sqrt{(2 \pi \sigma^2)^{D\cdot n}}} \exp\Big(-\frac{1}{2\sigma^2} \sum_{i=1}^{n} \norm{(x_i, y_i, z_i)^T - \bm{G}_i(\bm{\theta})}_2^2 \Big).
\end{equation}
The operator $\bm{G}$ outputs the RK4 solution of the Lorenz equations with respect to the input parameters $\bm{\theta}$, $D = 3$ is the dimension of the output,
and $n=30$ as above.

\begin{figure}[!htb]
    \centering
    \includegraphics[scale=1.1]{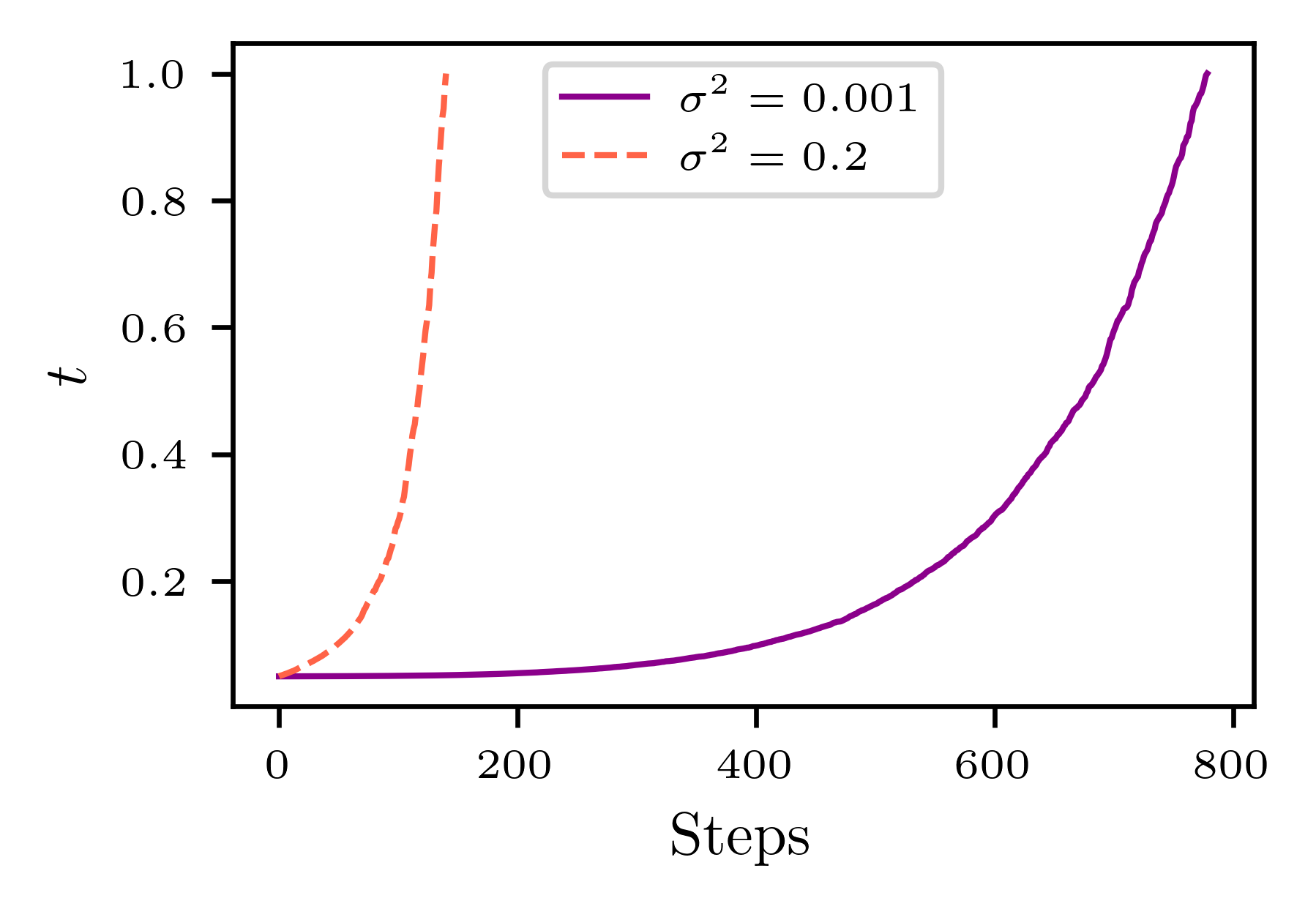}\vspace{-9pt}
    \caption{Annealing schedules from AdaAnn in example 4 (Lorenz attractor)}
    \label{fig:Lorenz_t_schedules}
\end{figure}

Starting with a base 
$q_{0}=\mathcal{N}(\mu = [10,10,10], \Sigma = 4 I_3)$,
we use planar flow with
$L = 250$ layers and apply the AdaAnn scheduler in Algorithm \ref{alg:adaptive} with the following hyperparameters: $\tau = 0.5$, $M=100$, $N=100$, $t_0=0.05$, $T_0=500$, $T=5$ and $T_1=\mbox{5,000}$. 
The learning rate for the Adam optimizer at $t<1$ is 0.0005; during the refinement phase at $t=1$, the batch size is increased to $N_1=200$ and a step learning rate scheduler is applied with a reduction of $\gamma=0.75$ every 500 training iterations.
The AdaAnn schedules are shown in Figure~\ref{fig:Lorenz_t_schedules}, which took 778 steps and 140 steps for $\sigma^2 = 0.001$ and $\sigma^2 = 0.2$, respectively.

\begin{figure}[!htb]
    \centering
    $\begin{array}{cc}
    \includegraphics[width=3in]{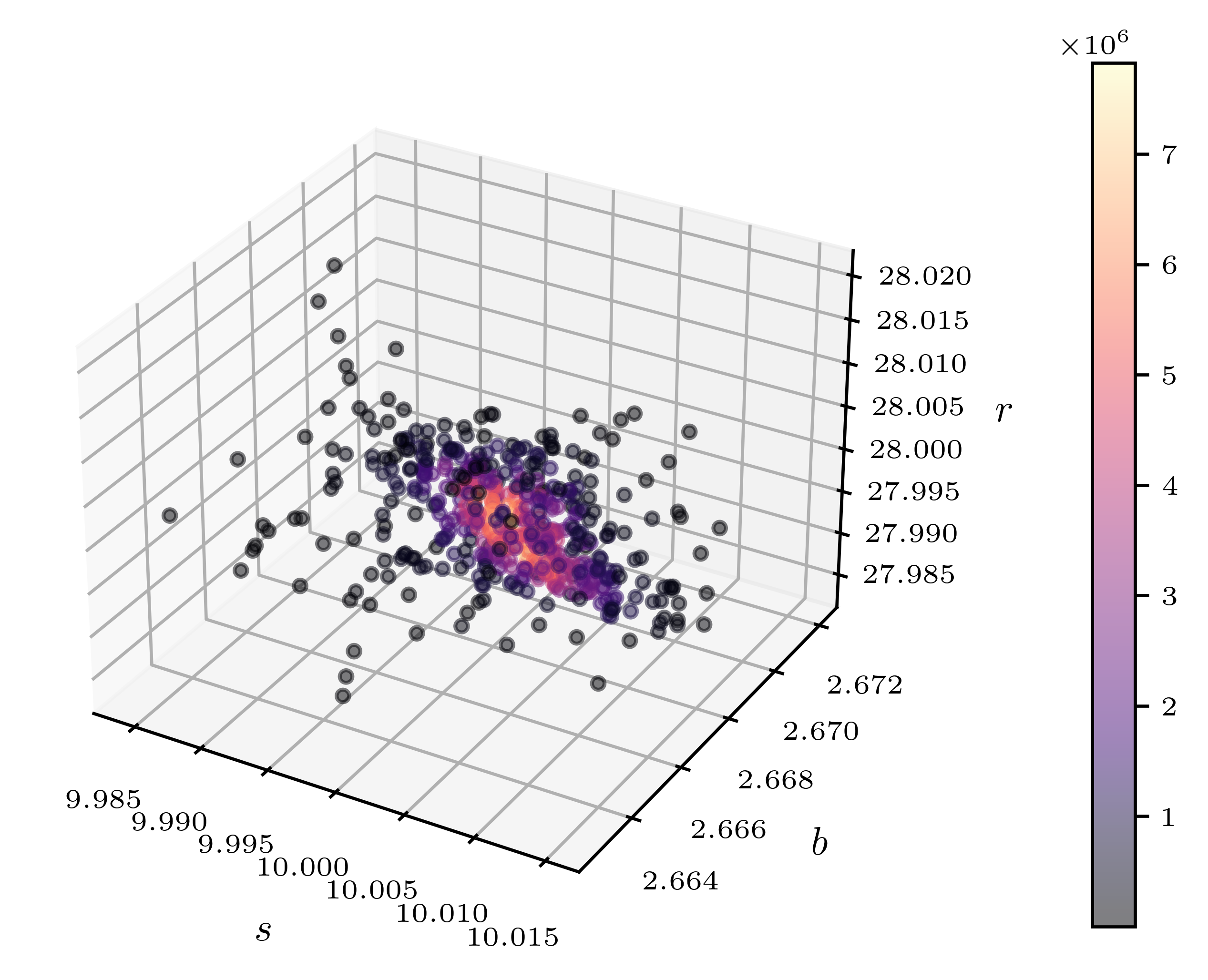} &
    \includegraphics[width=3in]{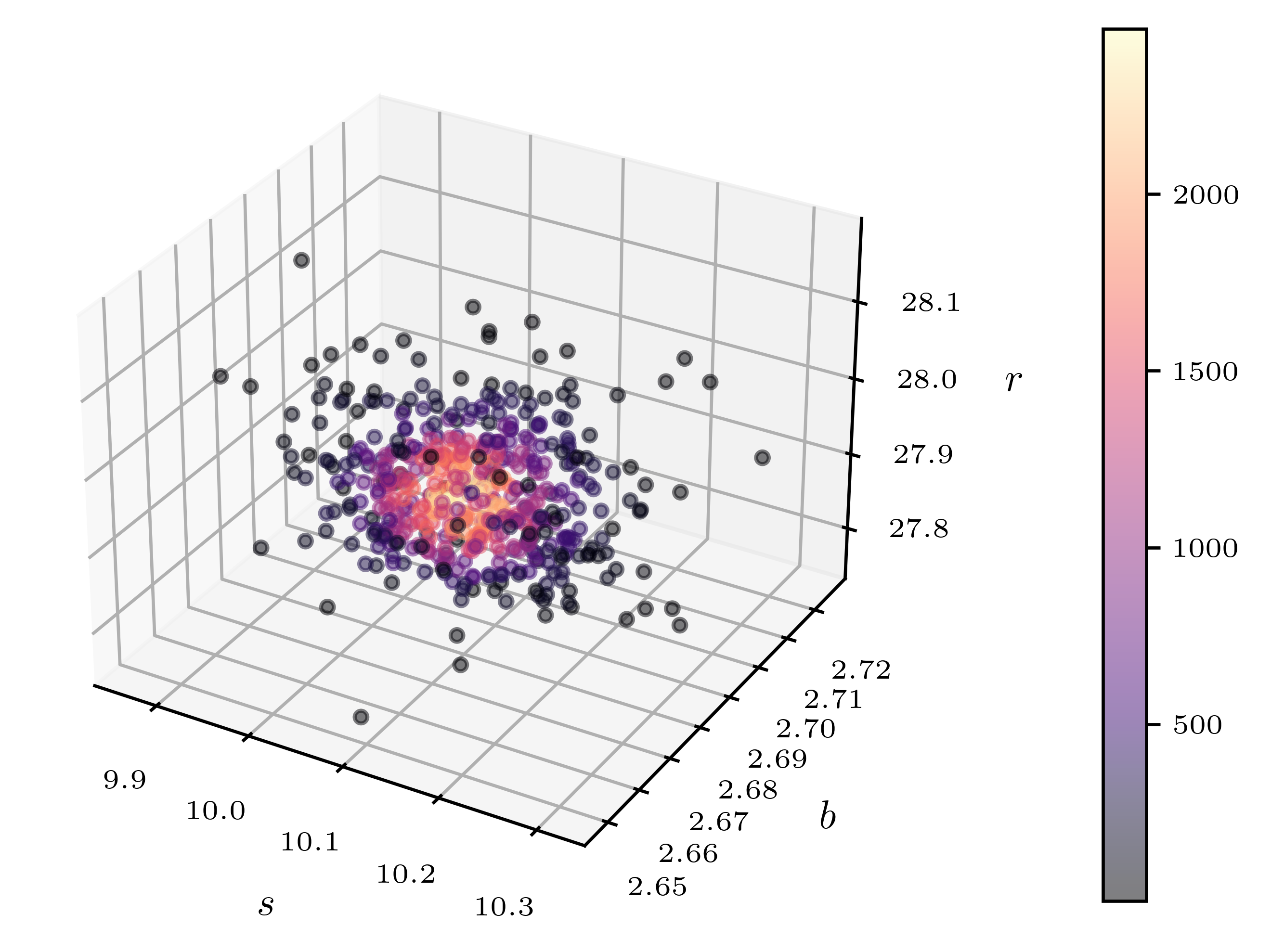} \\
    \hbox{(a) $\sigma^2 = 0.001$} & \hbox{(b) $\sigma^2 = 0.2$} \\
    \end{array}$
    \caption{Approximate posterior distribution of parameters $(s,b,r)$ for the Lorenz attractor obtained by VI-NF with AdaAnn}
    \label{fig:Lorenz_optimized}
\end{figure}

\begin{figure}[!htb]
\centering
{\includegraphics[scale=0.95]{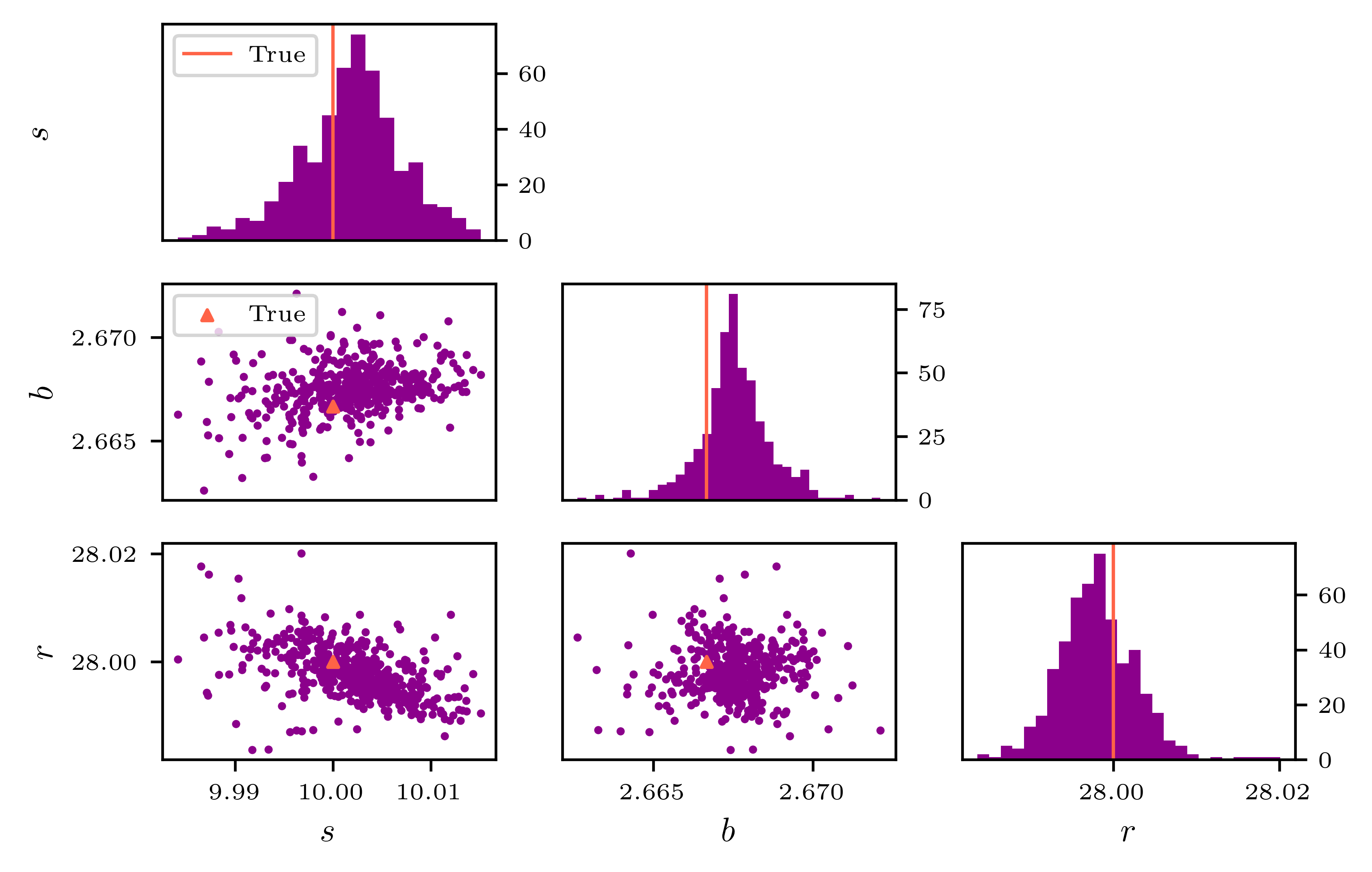}}\\
\vspace{-6pt}
(a) $\sigma^2 = 0.001$\\[8pt]
{\includegraphics[scale=0.95]{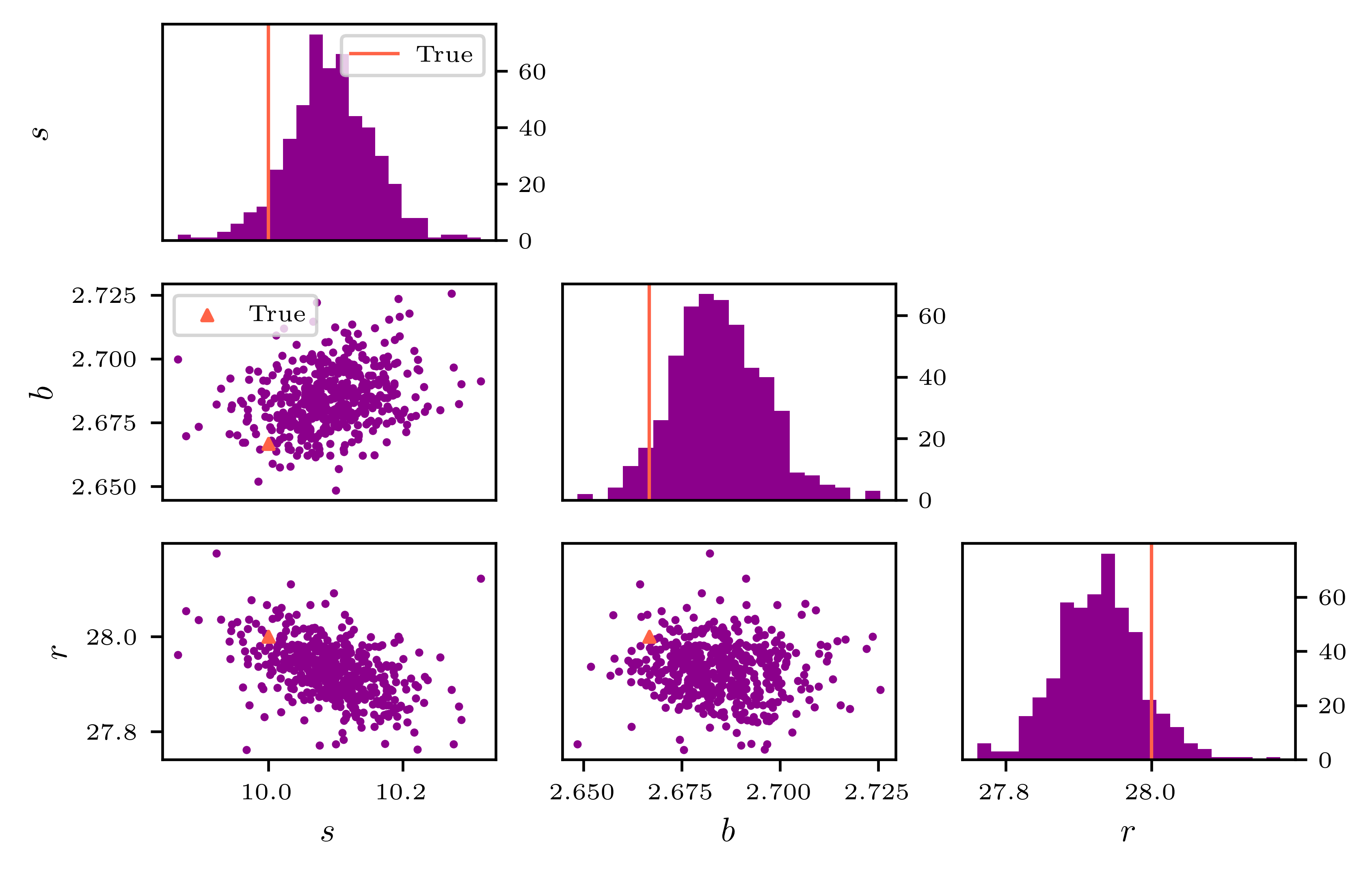}}\\
\vspace{-6pt}
(b) $\sigma^2 = 0.2$
\vspace{-6pt}
\caption{Marginal posterior distributions and pairwise scatter plots of parameters $(s,b,r)$ for the Lorenz attractor with $\sigma^2=0.001$ and $\sigma^2 = 0.2$.}
\label{fig:Lorenz_marginals}
\end{figure}

The resulting variational approximation $q_L(s,b,r|\bm X)$ is shown in Figure~\ref{fig:Lorenz_optimized}. The marginal histogram for each of the 3 parameters and the pairwise scatter plots are depicted in Figure~\ref{fig:Lorenz_marginals}. 
The inferred distributions agree well with the true parameter values. The MC estimates for the marginal means and standard deviations (SD) of the posterior distributions are computed from the final optimized approximate distribution $q_{L}$ using 10,000 samples and reported in Table~\ref{table:Lorenz_stats}. 
For each parameter, its true value is within one SD of the corresponding estimated parameter value.
\begin{table}[!htb]
\caption{Posterior mean and standard deviation for the parameters of the Lorenz attractor, based on 10,000 samples obtained by VI via NFs with AdaAnn.}\label{table:Lorenz_stats}
\centering
    \begin{tabular}[2in]{c cc c cc}
    \toprule
true &  \multicolumn{2}{c}{ $\sigma^2=0.001$} && \multicolumn{2}{c}{ $\sigma^2=0.2$} \\
     \cline{2-3}\cline{5-6}
    Parameter & posterior mean & posterior SD && posterior mean & posterior SD\\
    \midrule
    $s=10$ & 10.0021 & 0.0057 && 10.0925 & 0.0635 \\
    $b=8/3\approx2.6667$ & 2.6676 & 0.0012 && 2.6852 & 0.0125 \\
    $r=28$ & 27.9980 & 0.0055 && 27.9283 & 0.0557 \\
    \bottomrule
    \end{tabular}
\end{table}

\subsection{Example 5: ODE system for HIV dynamics}\label{sec:hiv}

This example infers the parameters of a system of ODEs that models the HIV dynamics \cite{hauenstein2019} based on the original system \cite{perelson2002}:
\begin{equation}\label{eq:HIV}
    \begin{aligned}
        \dot{x_1} &= p_1 - p_2 x_1 - p_3 x_1 x_3,\,\,\,\,\,\,\,\,\dot{x_2} = p_3 x_1 x_3 - p_4 x_2,\,\,\,\,\,\,\,\,\dot{x_3} = p_1 p_4 x_2 - p_5 x_3,\\
        y &= x_3.
    \end{aligned}
\end{equation}
In this system, $x_1$ is the number of $CD4^+$ T-cells that are susceptible to being infected by the HIV-1 virus and $x_2$ is the number of productively infected $CD4^+$ T-cells. The concentration of HIV-1 free virus, $x_3$, is measured in HIV-1 RNA per mL of plasma. 
The dynamics of the system are driven by the following five parameters: $p_1$ is the rate of target cells being produced from a source, $p_2$ is the rate of target cells dying, $p_3$ is the rate of target cells being infected by the HIV-1 virus, $p_4$ is the death rate of productively infected cells $x_2$, and $p_5$ is the clearance rate of infectious HIV-1 virus particles from the body.

We use VI via NFs to estimate the parameters $p_1$ and~$p_2$ along with the initial condition~$x_{2_0}$. The remaining parameters and initial conditions are considered known and fixed. The posterior distribution of $p_1$ and~$p_2$ may have a multimodal structure if this system has an identifiability degree greater than one \cite{hauenstein2019}. 
In fact, for this problem, the identifiability degree is~2 indicating two sets of parameter values producing an identical output, namely $\{p_1, p_2, x_{2_0} \}$ and $\{-p_1, p_2, -x_{2_0} \}$, generating the same observed trajectory on output~\mbox{$y(t)=x_3(t)$}.

\begin{figure}[!htb]
    \centering
    \includegraphics[scale=1.1]{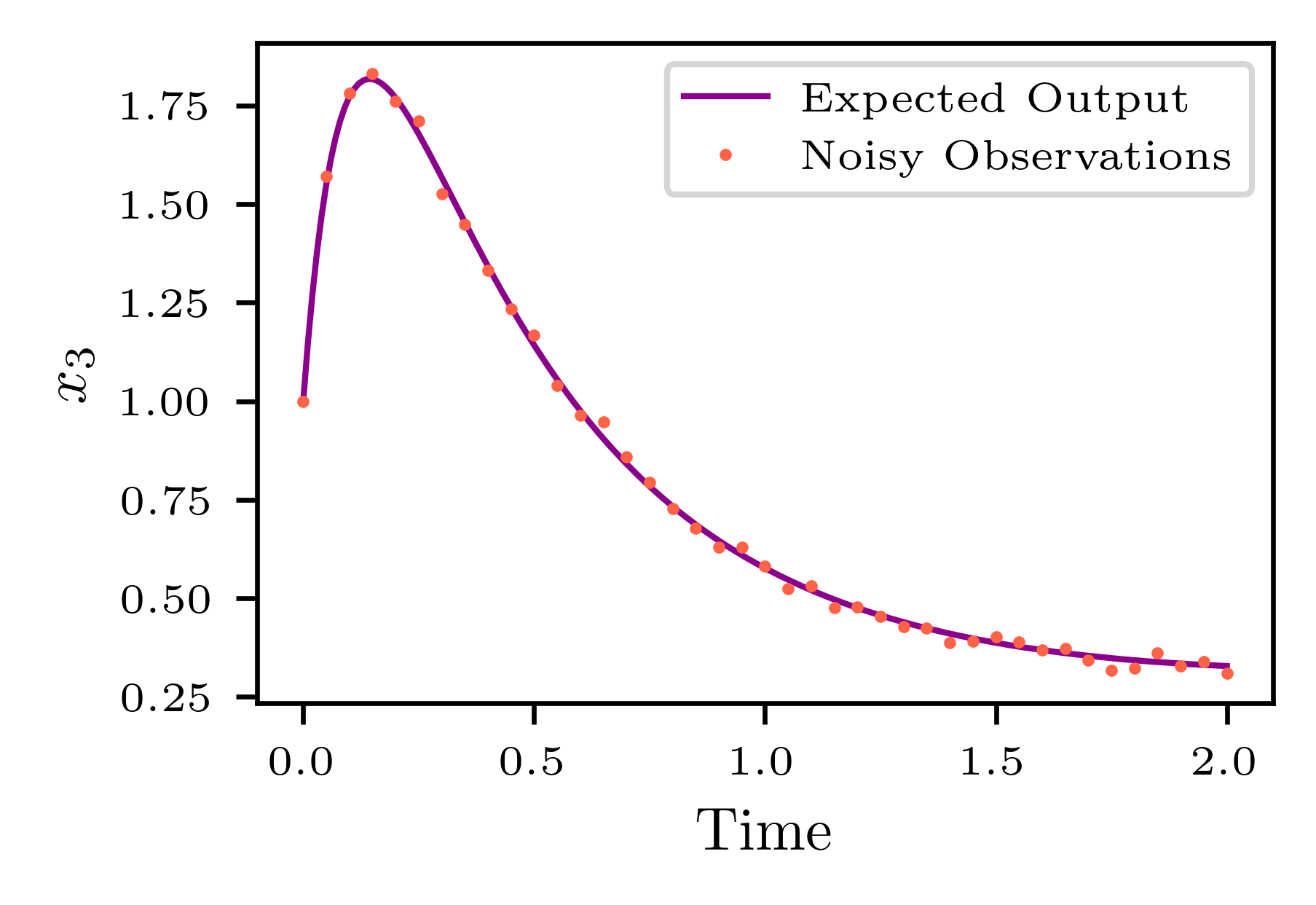}\vspace{-12pt}
    \caption{Trajectory of the output $x_3$ in units $10^4$~HIV RNA per mL of plasma over 2 months along with noisy data in example 5.}
    \label{fig:HIV_trajectory}\vspace{-6pt}
\end{figure}

The system~\eqref{eq:HIV} is numerically integrated using RK4 until time $t=2$ months with step size of $\Delta t=0.05$ months using the parameters and initial conditions in Table~\ref{table:HIV_parameterValues}. Synthetic data were generated using $n=40$ equally spaced data points from $y=x_3$ and adding 
Gaussian errors $\mathcal{N}(\mu=0, \sigma^2 = 0.0005)$ to the output solution $x_3$, as shown in Figure~\ref{fig:HIV_trajectory}.
\begin{table}[!htb]
\centering
    \begin{tabular}[2in]{ccc}
    \toprule
    \bf{Unknown Parameters} & \bf{Known Parameters} & \bf{Fixed Initial Conditions} \\
    \midrule
    $p_1 = 1.2$ & $p_3 = 4.1$ & $x_{1_0} = 0$ \\
    $p_2 = 0.8$ & $p_4 = 10.2$ & $x_{3_0} = 1$ \\
    $x_{2_0} = 1.5$ & $p_5 = 2.6$ & \\
    \bottomrule
    \end{tabular}
    \caption{Parameter values and initial conditions in the HIV dynamics ODE system.}
    \label{table:HIV_parameterValues}
\end{table}

The posterior distribution of parameters $\bm\theta=\{p_1, p_2, x_{2_0}\}$ given $n$ observed data points on $x_3$ is 
$p(\bm\theta|\bm x_3) \propto (2 \pi \sigma^2)^{-n/2} \exp\big(\sigma^{-2} \sum_{i=1}^{n} {(x_{3i} - G_i(\bm{\theta})})^2 \big)$
given the  Gaussian likelihood function and  a uniform prior on $\bm\theta$.
To approximate this posterior, we transform a base distribution $\mathcal{N}(\mu=[0,0,0], \Sigma=4I_3)$ using a composition of $L=250$ planar flows with hyperbolic tangent activation functions. 
We run AdaAnn with $t_0=0.00005$, $\tau=0.005$,  $T_0=\mbox{1,000}$, $T=5$, $N=100$, and $M=100$. The learning rate for the Adam optimizer is 0.0005. Once we reach $t=1$, we refine the posterior approximation by training for an additional $T_1=\mbox{5,000}$ iterations, increasing the batch size to $N_1=200$, and adopting a step learning rate scheduler for the Adam optimizer (with learning rate reduced by a factor of $\gamma=0.75$ after 1,000 training iterations).  

\begin{figure}[!htb]
    \centering
    \includegraphics{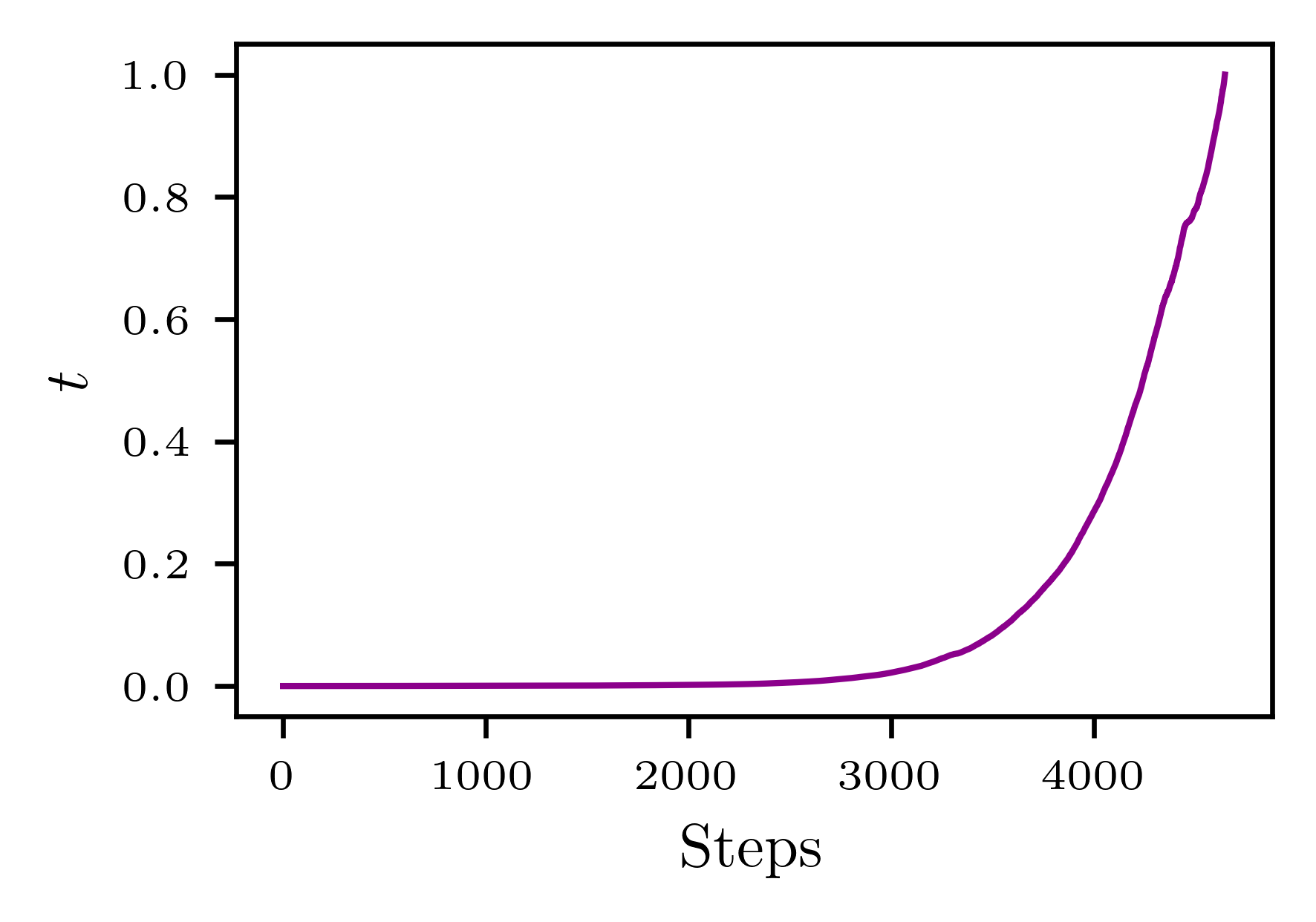}\vspace{-12pt}
    \caption{AdaAnn schedule in Example 5 (the HIV dynamics ODE system)}
    \label{fig:HIV_schedule}
\end{figure}

\begin{figure}[!htb]
    \centering
    $\begin{array}{cc}
    \includegraphics[width=3in]{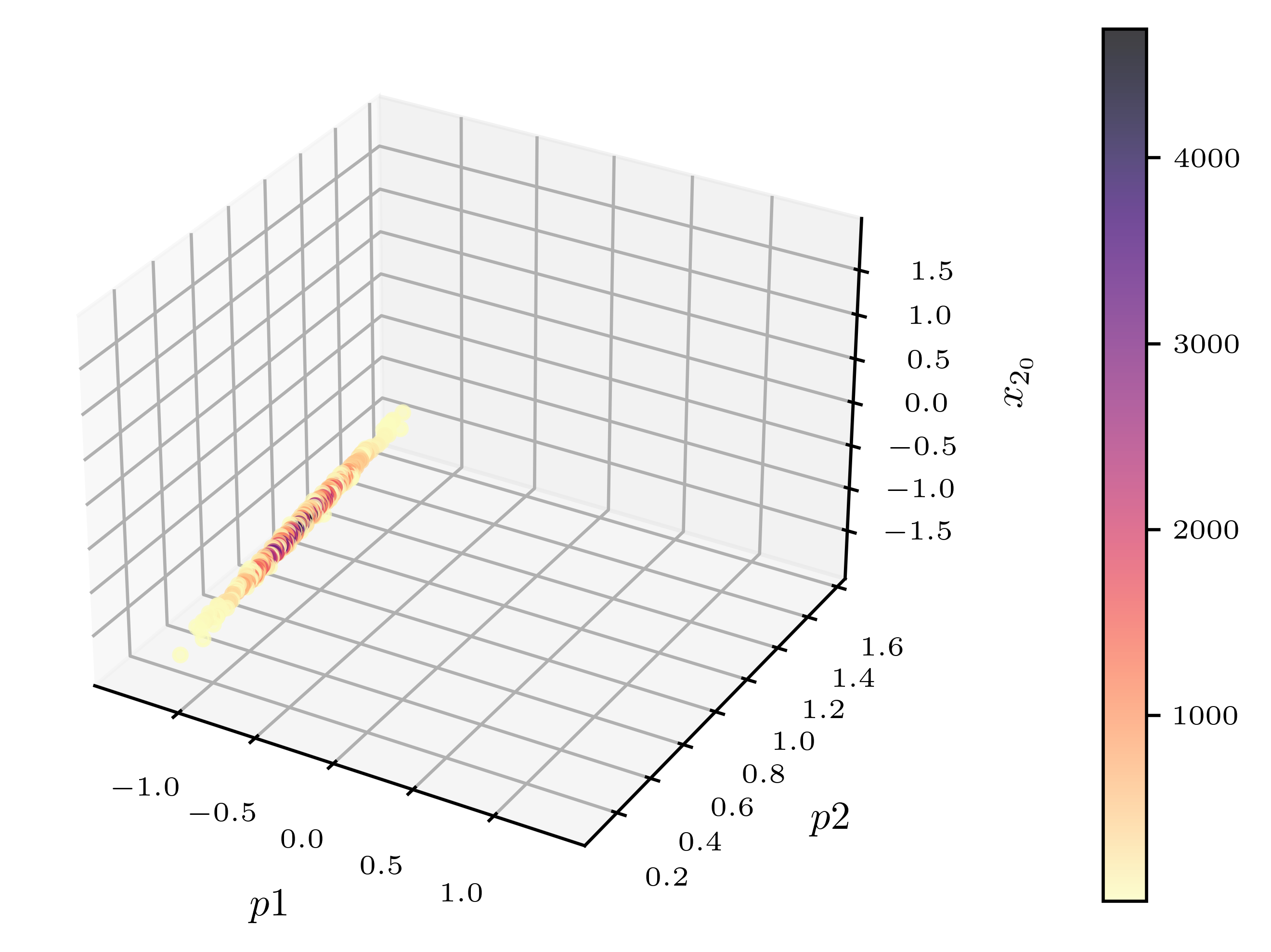} &
    \includegraphics[width=3in]{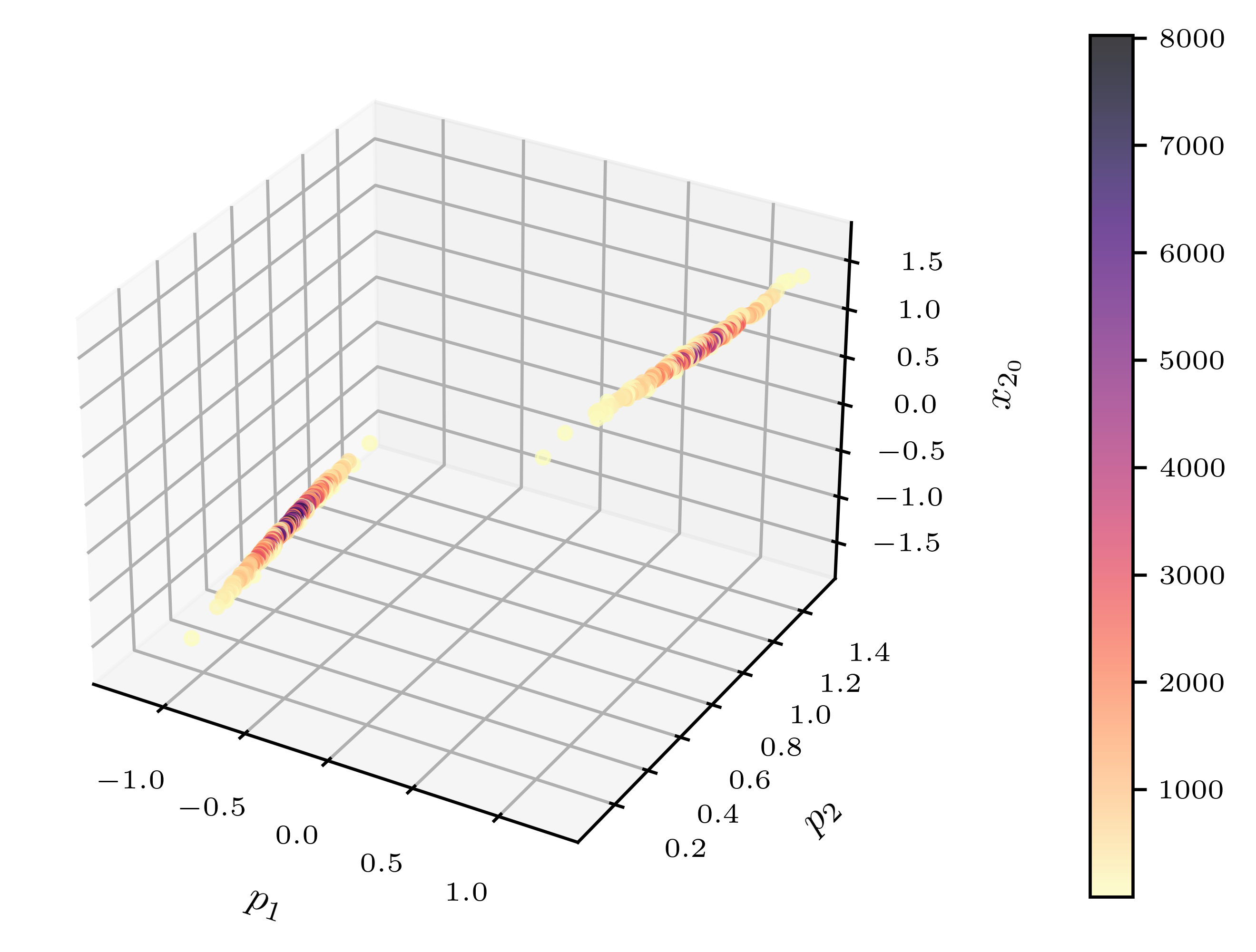} \\
    \hbox{(a) No-Annealing} & \hbox{(b) AdaAnn}
    \end{array}$
    \caption{Approximate posterior distributions without annealing versus with AdaAnn in Example 5}
    \label{fig:HIV_approxDistributions}
\end{figure}

The AdaAnn schedule is depicted in Figure~\ref{fig:HIV_schedule} with a total of 4,645 steps. The resulting approximation $q_L$ captures the bimodal structure of the target posterior distribution, as presented in Figure~\ref{fig:HIV_approxDistributions}. For comparison, we also run the planar flow without annealing for 20,000 iteration. The resulting $q_L$ inconsistently converges to either a unimodal or bimodal approximation. 
Since only the mode with positive parameters is biologically relevant, converging to one with negative parameters may lead to the conclusion that the model is unable to reproduce the observed behavior with physically sound parameters.
%

\begin{figure}[!htb]
    \centering
    \includegraphics{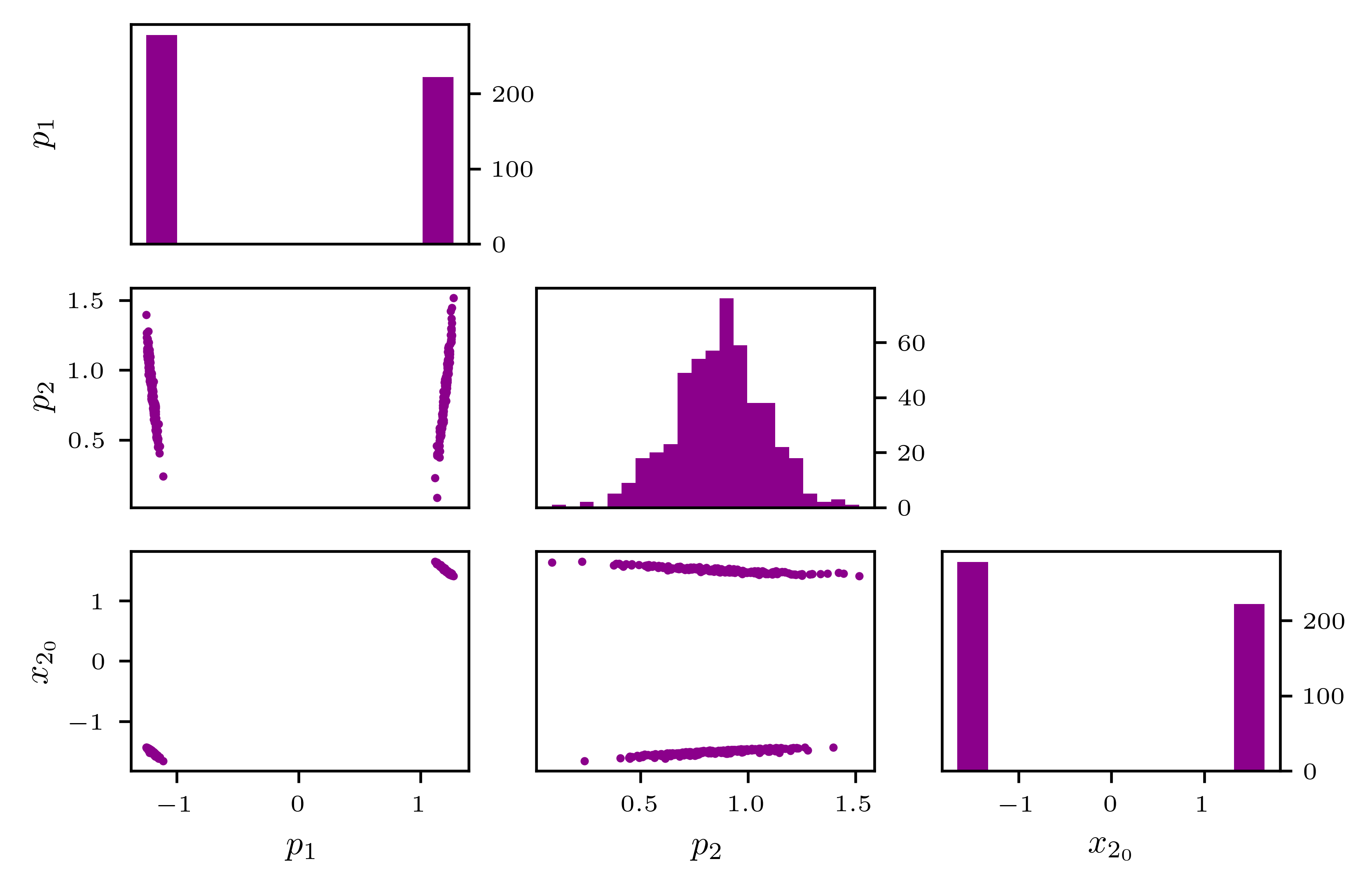}
    \caption{Marginal distributions for the HIV dynamics ODE system.}
    \label{fig:HIV_marginals}
\end{figure}

\begin{figure}[!htb]
    \centering
    \includegraphics{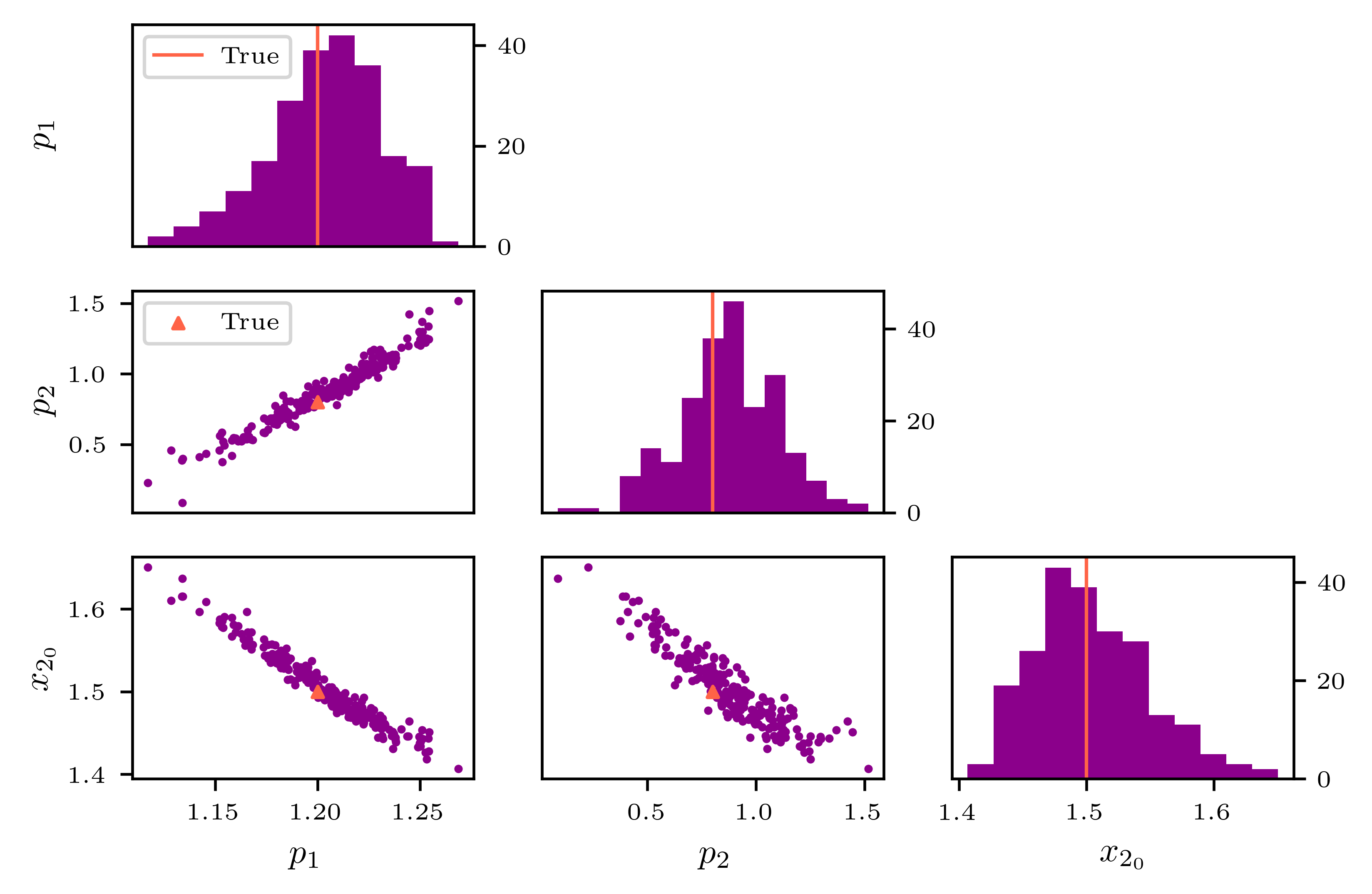}
    \caption{Marginal distribution of the biologically admissible mode for the HIV dynamics ODE system.}
    \label{fig:HIV_marginals_singleMode}
\end{figure}

The marginal distributions are also shown in Figure~\ref{fig:HIV_marginals}. 
Since the left mode is not biologically meaningful due to negative parameter values, we also included the marginal distributions for the right mode plotted against the true parameter values in Figure~\ref{fig:HIV_marginals_singleMode}. 
The true model parameters are accurately inferred by combining VI and NFs with the proposed adaptive annealing schedule. The posterior marginal means and standard deviations are computed using 10,000 samples and displayed in Table~\ref{table:HIV_stats}. 
\begin{table}[!htb]
\centering
    \begin{tabular}[2in]{cccccc}
    \toprule
    true parameter & \multicolumn{2}{c}{Biologically unadmissible} && \multicolumn{2}{c}{Biologically admissible} \\
    \cline{2-3}  \cline{5-6}
    & Posterior mean & Posterior SD  && Posterior Mean & Posterior SD \\
    \midrule
    $p_1=1.2$ & -1.2014 & 0.0249 & & 1.2019 & 0.0274 \\
    $p_2=0.8$ & \,\,0.8485 & 0.2001 & & 0.8609 & 0.2196 \\
    $x_{2_0}=1.5$ & -1.5067 & 0.0412 &  & 1.5060 & 0.0452 \\
    \bottomrule
    \end{tabular}
    \caption{Posterior mean and standard deviation of the parameters for the HIV dynamics ODE system (Example 5).}
    \label{table:HIV_stats}
\end{table}

\subsection{Summary of the Examples}

\begin{figure}[!hbt]
    \centering\vspace{-6pt}
    \includegraphics[width=3.5in]{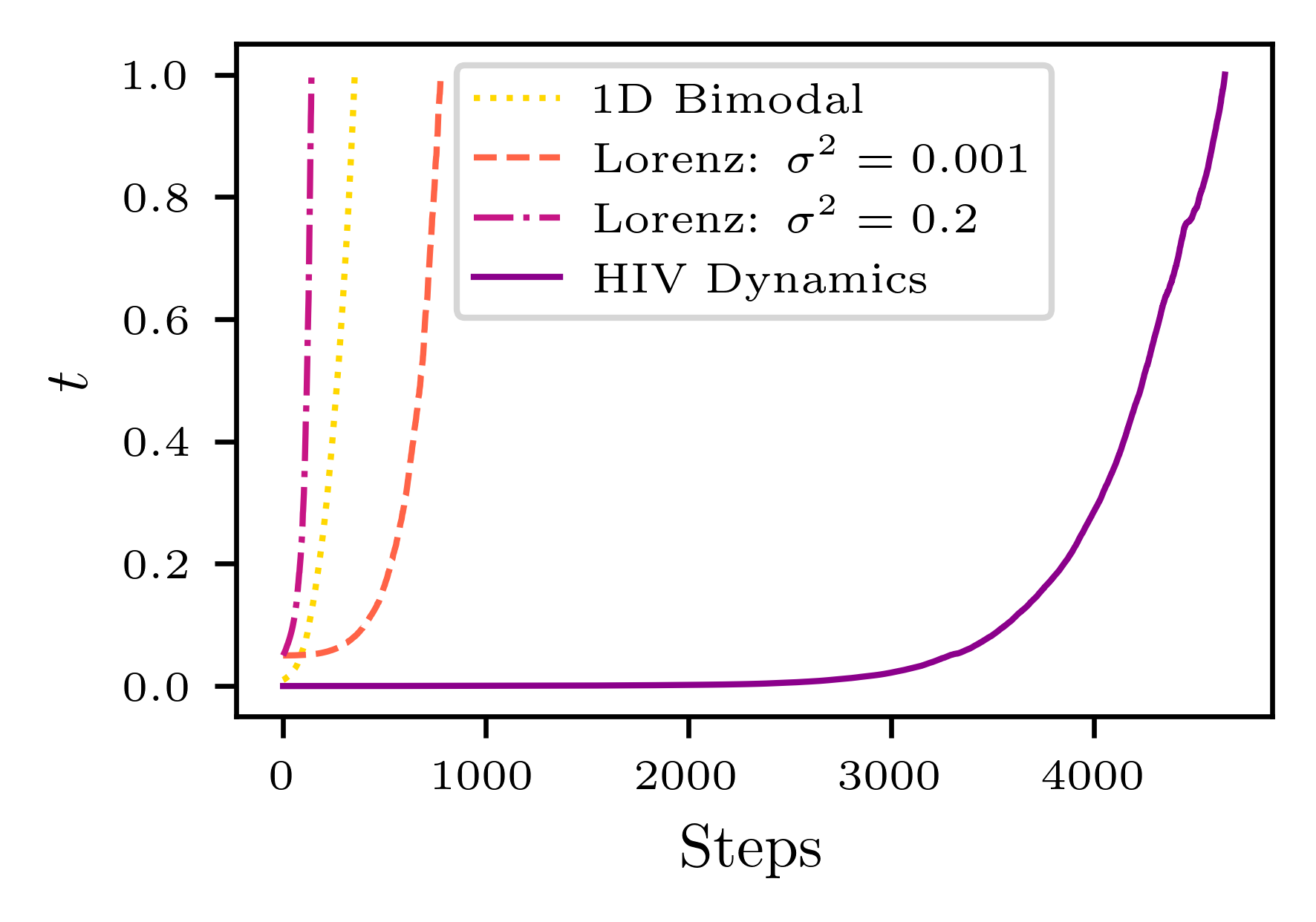}\vspace{-6pt}
    \caption{Comparison of annealing schedules for one-dimensional bimodal density (example 1), the Lorenz attractor (example 4), and HIV dynamical system (example 5).}
    \label{fig:schedules_comparison}\vspace{-3pt}
\end{figure}

The target distributions in these fives examples are of varying degrees of complexity and AdaAnn produces distinct annealing schedules that are well adapted to the complexity of the underlying posterior distribution. This is evident from Figure~\ref{fig:schedules_comparison} that illustrates the evolution of the inverse temperature generated by AdaAnn for Examples 1, 4 and 5. 
A fast schedule with a limited number of small temperature increments is produced for the approximation of the one-dimensional bimodal density. 
A larger noise variance in the data for the Lorenz system (i.e., $\sigma^2=0.2$) leads to a wider posterior distribution that AdaAnn is able to approximate in few, mainly large, steps. A reduced variance ($\sigma^2=0.001$) corresponds instead to a more sharply peaked posterior which requires more small increments near the beginning.
For the bimodal HIV dynamics posterior in 3D, characterized by two well separated peaks, AdaAnn requires significantly more steps and a smaller initial temperature, as expected.
It is also interesting to observe that, in the schedule for the HIV dynamical system example, $\epsilon_{k}$ is reduced after $\sim$ 4,500 iteration, producing a small but visible ``kink'' in the temperature schedule and it appears consistently in multiple 
runs. Further investigation is needed to better understand this phenomenon and what features of the target distribution or the approximate distribution at $t$ causes the annealing process
to slow down.

The relevant hyperparameters for AdaAnn (Algorithm \ref{alg:adaptive}) in the five examples are summarized in Table~\ref{table:hyperparameters}.
\begin{table}[!htb]
\centering
    \resizebox{\textwidth}{!}{
    \begin{tabular}[2in]{cccccccccc}
    \toprule
    \bf{Example} & \bf{Description} & $\bm{\tau}$ & $\bm{t_{0}}$ & $\bm{T_0}$ & $\bm{T}$ & $\bm{T_1}$ & $\bm{N}$ & $\bm{N_1}$ & $\bm{M}$ \\
    \midrule 
    1 & 1D bimodal & 0.01 & 0.01 & 500 & 2 & 8,000 & 100 & 1,000 & 1,000 \\ 
    2 & 1D Parametric bimodal & 0.005 & 0.01 & 500 & 5 & - & 100 & - & 1,000 \\ 
    3 & 2D bimodal density & 0.01 & 0.01 & 500 & 5 & - & 100 & - & 1,000 \\
    4 & Lorenz attractor & 0.5 & 0.05 & 500 & 5 & 5,000 & 100 & 200 & 100 \\
    5 & HIV model & 0.005 & 0.00005 & 1,000 & 5 & 5,000 & 100 & 200 & 100 \\ 
    \bottomrule
    \end{tabular}}
    \caption{Summary of the AdaAnn hyperparameters used in all 5  examples.}
    \label{table:hyperparameters}
\end{table}
One may also want to allow for more gradient updates to be performed for each $t_{k}$ so that NFs can provide a better approximation of $p^{t_{k}}(\bm{Z},\bm{X})$, especially for more complex or higher dimensional densities. Except for the motivating example in Section~\ref{sec:Comparing_Methods} where we perform two gradient updates per temperature increase (linear scheduler), we use~5 updates in all of the other examples. 
At the target temperature of $t=1$, it is also desirable to perform additional iterations to refine the approximation of the target distribution. For the Lorenz and HIV dynamical system, 5,000 appears to be a reasonable number of iterations leading to an accurate posterior. At $t=1$, we also typically increase the the batch size (e.g., Examples 1, 4 and 5).

\section{Discussion}\label{sec:discussion}

We introduced AdaAnn, an adaptive scheduler that automatically suggests changes in the annealing temperature when using NFs for VI. This scheme has third-order accuracy and is obtained from a Taylor series expansion of the KL divergence between two annealed densities which differ by a sufficiently small inverse temperature increment. 

AdaAnn requires two main parameters to be defined: the initial temperature $t^{-1}_0$ and the KL divergence tolerance $\tau$. The choice of $t_0$ is dependent on the separation and width of the modes in the target distribution. 
As observed for the HIV dynamical system in Section~\ref{sec:hiv}, a posterior with very narrow or separated modes requires a smaller $t_0$, leading to a more uniform initial density.
Regarding the KL divergence tolerance, an exceedingly large~$\tau$ can provide a poor approximation that misses relevant features in the target distributions, e.g., could miss 
one of the modes in a multimodal posterior. Conversely, a too small $\tau$ may result in unnecessary incremental steps and added computational cost yielding no edge in computational efficiency 
over linear~schedulers. 

AdaAnn is simple to implement and can lead to significant computational saving compared to {\em a priori} selected annealing schedules. We demonstrate the application of AdaAnn in planar flows for distribution approximation and variational inference, but no problem is foreseen in applying AdaAnn with other types of flows or other algorithms for the solution of inverse problems (e.g., MCMC).

\section*{Acknowledgements}
All authors gratefully acknowledge the support 
by the NSF Big Data Science \& Engineering grant \#1918692
and the computational resources provided through the Center for
Research Computing at the University of Notre Dame.
DES also acknowledges support from
NSF CAREER grant \#1942662.

\bibliographystyle{abbrv}
\bibliography{PlanarFlowBib}

\end{document}

%% file: Preamble/preamble.tex
\usepackage[margin=1in]{geometry}

\usepackage{amsmath}
\usepackage{amssymb}
\usepackage{amsfonts}
\usepackage{amsthm}
\usepackage{bm}
\usepackage{listings}
\usepackage{mathtools}
\usepackage{physics, comment}

\usepackage{authblk}

\usepackage{graphicx}
\usepackage{float}

\usepackage{booktabs}
\usepackage{multirow}
\usepackage{siunitx}

\theoremstyle{plain}
\newtheorem{theorem}{Theorem}

\usepackage{xcolor}

\usepackage[colorlinks=true,linkcolor=blue,citecolor=blue]{hyperref}

\usepackage{algorithm}
\usepackage{algpseudocode}